\documentclass{aastex631}
\shorttitle{Distorted model for flux ropes}
\shortauthors{Nieves-Chinchilla et al.}
\graphicspath{{./}{figures/}}

\begin{document}

\title{Distorted-Toroidal Flux Rope model for Heliospheric Flux Ropes}
\shortauthors{Nieves-Chinchilla, Hidalgo and Cremades}

\author[0000-0003-0565-4890]{Teresa Nieves-Chinchilla}
\affiliation{Heliospheric Physics Lab. Heliophysics Science Division, NASA- Goddard Space Flight Center. \\
8800 Greenbelt Rd., \\
Greenbelt, MD 20770, USA}

\author[0000-0003-1617-2037]{Miguel Angel Hidalgo}
\affiliation{Department of Physics and Mathematics, University of Alcalá\\
Madrid, Spain}

\author[0000-0001-7080-2664]{Hebe Cremades}
\affiliation{Grupo de Estudios en Heliofísica de Mendoza, CONICET, Universidad de Mendoza\\ 
665 Boulogne Sur Mer, Mendoza 5500, Argentina}



\begin{abstract}
The three-dimensional characterization of magnetic flux-ropes observed in the heliosphere has been a challenging task for decades. This is mainly due to the limitation to infer the 3D global topology and the physical properties from the 1D time series from any spacecraft. To advance our understanding of magnetic flux-ropes whose configuration departs from the typical stiff geometries, here we present the analytical solution for a 3D flux-rope model with an arbitrary cross-section and a toroidal global shape. This constitutes the next level of complexity following the elliptic-cylindrical (EC) geometry. The mathematical framework was established by \citeauthor{Nieves-Chinchilla2018ApJ}, with the EC flux-rope model that describes the magnetic topology with elliptical cross-section as a first approach to changes in the cross-section. In the distorted-toroidal flux rope model, the cross-section is described by a general function. The model is completely described by a non-orthogonal geometry and the Maxwell equations can be consistently solved to obtain the magnetic field and relevant physical quantities. As a proof of concept, this model is generalized in terms of the radial dependence of current density components. The last part of this paper is dedicated to a specific function, $F(\varphi)=\delta(1-\lambda\cos\varphi)$, to illustrate possibilities of the model. This model paves the way to investigate complex distortions of the magnetic structures in the solar wind. Future investigations will in-depth explore these distortions by analyzing specific events, the implications in the physical quantities, such as magnetic fluxes, heliciy or energy, and evaluating the force balance with the ambient solar wind that allows such distortions.    

\end{abstract}



\section{Introduction} 
\label{sec:intro}
In heliophysics, a flux rope could be defined as a magnetized plasma confined within magnetic field lines wrapping around an axis that transports mass, magnetic flux, energy and helicity away from the Sun. In an effort to create an unidealized picture of a flux rope in the heliosphere, it could be described by an internal complex current density distribution driving a twisted but not necessarily ordered magnetic field topology that maintains the plasma enclosed. In this description, the global geometry would be determined by the flux rope genesis back at the Sun as well as by the dynamical balance with the ambient solar wind as the flux rope evolves in the heliosphere. 

Little is known about the internal structure and global shape of heliospheric flux ropes, basically because of the lack of observations that are limited to a few locations in the heliosphere. The progressive increase of space-based telescopes, such as Parker Solar Probe \citep[PSP,][]{Fox2016} and Solar Orbiter \citep{Muller2020} that sum up to SOHO \citep[][]{domingo1995}, STEREO \citep[][]{Kaiser2008}, and SDO \citep[][]{Pesnell-etal2012}, are providing a valuable combination of remote-sensing observations that partially enable to untangle the third dimension from the ecliptic-based field of view. 

On the basis of in-situ observations, the magnetic flux ropes observed in the heliosphere have been assumed for decades to be a force-free magnetic structure in a simple circular-cylindrical geometry \citep{Lundquist1951,Burlaga1981,Suess1988,Lepping1990} but also exploring the magnetohydrostatic force balance \citep{Hidalgo2002,Hidalgo2002a,Sonnerup1996,HauSonnerup1999,Hu2017}. 
In an effort to reconcile this view with the remote-sensing observations, a realistic flux rope may depart from that idealized picture as it propagates through the corona and in the interplanetary medium. 
The current understanding of the observed patterns in these observations 
suggests that the dark, round void outlined by excess brightness is the flux rope cavity with its axis seen oriented along the line of sight, while the bright front defines the leading edge associated with part of the sheath in interplanetary in situ data \citep[e.g.][]{Rouillard2011,Kilpua2017}.

Definitively, the upcoming out-of-the-ecliptic observations from Solar Orbiter are promising towards unraveling the global shape of large structures in the inner heliosphere. In parallel to the increase of available observations, the understanding of the fundamental physics associated with flux ropes also requires models adapted to the complexity of the space environment. We approach this challenge with the revision of the circular-cylindrical model in \citet[][henceforth CC model]{Nieves-Chinchilla2016}, which provided complexity in the flux rope magnetic structure by including the polynomial series in the current density. In a second paper, \citet[][henceforth EC model]{Nieves-Chinchilla2018ApJ}, we developed the mathematical formulation to solve any magnetohydrodynamic equations in a non-orthogonal coordinate system, and approached the geometrical complexity with an elliptical cross section for the cylinder as an approximation to a distorted flux rope. 

This paper aims to advance in the development of a model that better converges to the above definition of a heliospheric flux rope, namely that described by an internal complex current density distribution driving a twisted but not necessarily ordered magnetic field topology. To that aim, we develop a 3D flux rope model based on a toroid but allowing more complex cross-section geometries. Section~2 includes the mathematical details of a general model and the physical quantities such as magnetic fluxes, energy, helicity and Lorentz force. In Section~3, we adapt the model to a specific cross-section and compare with the imaging observations with the cylindrical case. Section~4 includes a brief discussion and final remarks.

\section{General Distorted-Toroidal Flux Rope Model}
\label{Sec:CoordSys} 

Following the path of the previous papers and assuming a toroidal-shaped geometry, we introduce a distorted-toroidal (DT) coordinate system here, 
\begin{eqnarray}
\label{Eq:CoordSyst}
x &=&[\rho+ rF\cos\varphi]\cos\psi \nonumber \\
y &=&[\rho+ rF\cos\varphi]\sin\psi \\
z &=&  r\sin\varphi \nonumber 
\end{eqnarray}
where $\rho$ is the major radius for the torus, $\varphi$ and $\psi$ are the poloidal and toroidal angles, and $F=F(\varphi)$ is a function that geometrically characterizes the cross-section distortion.   

\begin{figure}
    \centering
   \includegraphics[width=0.89\textwidth]{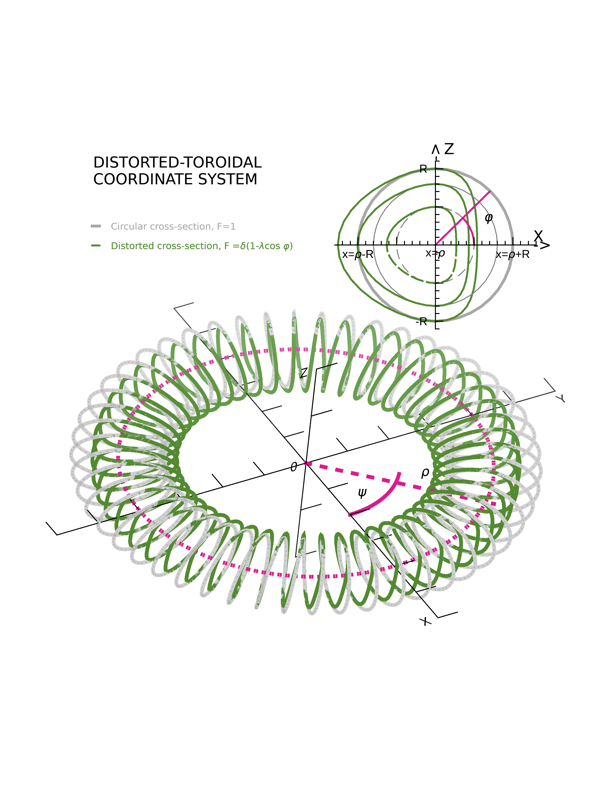}
    \caption{Representation of a circular-toroidal flux-rope (gray line) and a distorted flux rope with an uniform-twist field. In pink color the  $r,\psi,\varphi$ - components, $\rho$ is the major radius, $R$ minor radius, $\psi$ is the azimuth angle, and $\varphi$ the poloidal angle. Both structures are represented in the same distorted-toroidal coordinate system with F=1 (gray color) for the circular case and the distorted case (green color) with F=$\delta(1-\lambda cos\varphi)$, $\rho =4$, $\delta=0.9$ and $\lambda$=0.4.  }
    \label{fig:figure1}
\end{figure}

Figure~\ref{fig:figure1} illustrates the global 3D global geometry and a detail of the distorted (green color) and non-distorted (gray color) cross-sections of a uniform-twist magnetic flux-rope based on this coordinate system. The new coordinates are indicated in pink color in the graphics. The $\psi-$ and $\varphi-$ angles range from $0-2\pi$ and the $r-$coordinate ranges from 0 to R. The $r$-coordinate is not the distance to the center of the flux-rope that will be defined by the $rF(\varphi)$ function and will shape the cross-section. In the case of Figure~\ref{fig:figure1}, we have selected the function $F=\delta (1-\lambda\cos\varphi)$ with $\delta$= 0.9 and $\lambda$=0.4. In this case, the structure is highly compressed in the outer edge of the torus but it extends beyond the $r=R$ value at the inner edge of the torus. The amount of front compression or rear extension in this case is determined by the two parameters, $\delta$ and $\lambda$. 
In the more general case, the $F$ function may be generalized to any angular function, depending on the poloidal angle $\varphi$ and on the number of parameters needed to characterize the cross-section. 

\begin{figure}
    \centering
   \includegraphics[width=0.72\textwidth]{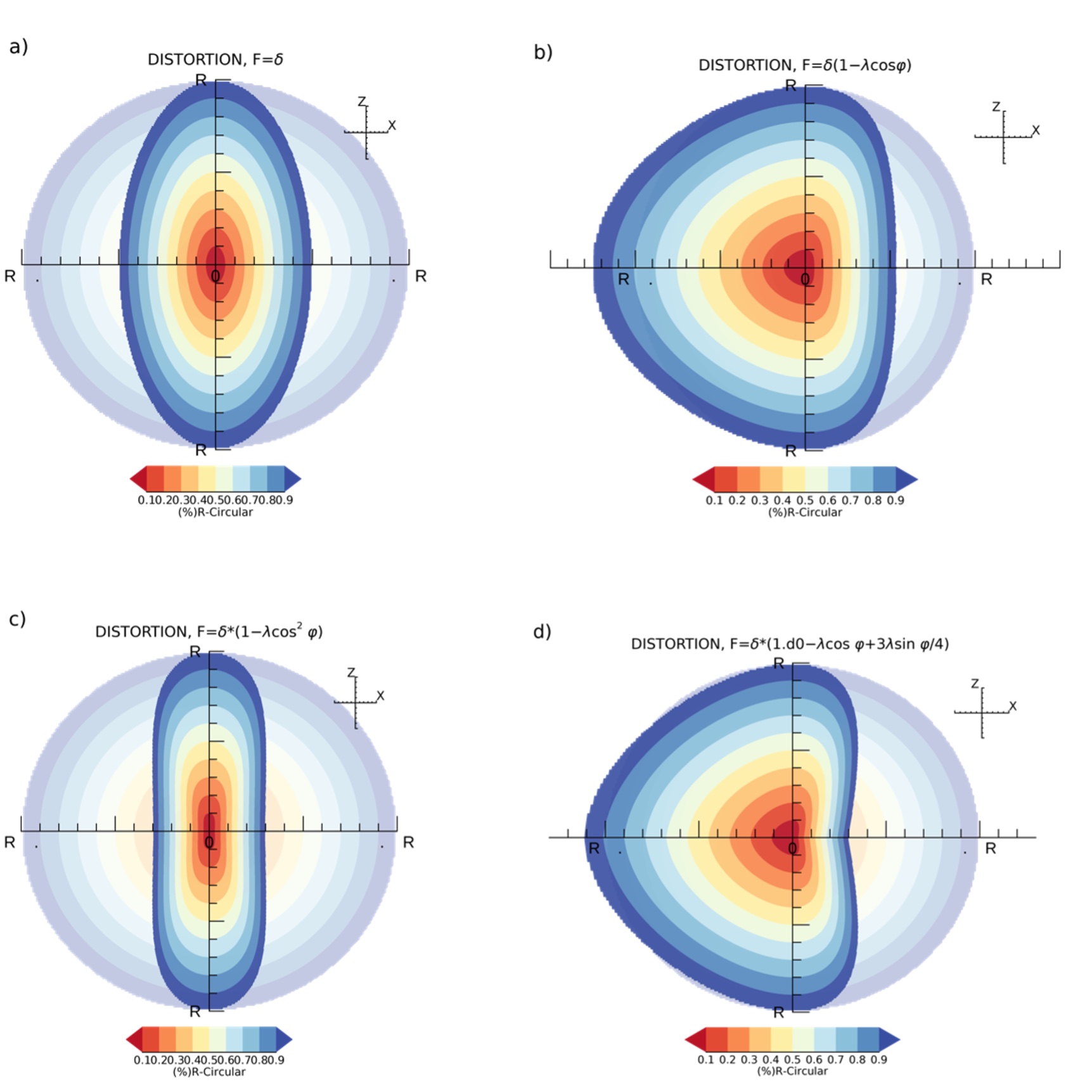}
    \caption{Illustration of different cross-sections based on (a) F=$\delta$, elliptical cross-section, (b) $F=\delta(1-\lambda cos\varphi)$ with $\delta=0.9$ and $\lambda$=0.4, (c) $F=\delta(1-\lambda\cos^2\varphi)$ with $\delta=0.5$ and $\lambda$=0.4, and (d) $F=\delta(1-\lambda\cos\varphi+3\lambda\sin\varphi/4)$ with $\delta=0.9$ and $\lambda$=0.4. In all simulations, $\rho=3R$. Each case is displayed on top of the semi-transparent circular cross-section to highlight the distortion.   }
    \label{fig:figure2}
\end{figure}

Figure~\ref{fig:figure2} illustrates some other  possibilities of the DT coordinate system for four $F$-functions. In the four cases, the colored cross-sections are shown in contrast with a semitransparent circular cross-section ($F=1$). Figures~\ref{fig:figure2}a,b illustrate the previously discussed cases for $F=\delta$ and $F=\delta(1 - \lambda\cos\varphi)$. In this last case, the change in the sign in the $\lambda$ parameter will determine the distortion face, outwards for the negative sign, inwards for the positive sign. The other two examples, Figure~\ref{fig:figure2}c and d exemplify other two observed shapes that may remind some white light observations of CMEs in the heliosphere.  

\subsection{Model framework}
Since the DT coordinate system is necessarily  orthogonal, any physical quantity should be described by a set of the \textit{covariant} and \textit{contravariant} components. Here, we will follow the methodology described in \citet{Nieves-Chinchilla2018ApJ}, just providing the critical description for this paper. 

The basis vectors will be defined as
\begin{eqnarray}
\label{Eq:BsVectors}
\vec{\varepsilon}_{r} &=&
    [F\cos\varphi\cos\psi,
    F\cos\varphi\sin\psi,
    \sin\varphi]  \nonumber \\
\vec{\varepsilon}_{\psi} &=&
    [ -[\rho+rF\cos\varphi]\sin\psi, 
    [\rho+rF\cos\varphi] \cos\psi, 0]  \\
\vec{\varepsilon}_{\varphi}  &=&
     r[ 
    \Im\cos\psi ,
    \Im\sin\psi,
    \cos\varphi] \nonumber 
\end{eqnarray}
where 
\begin{eqnarray}
    \Im &=& \Im(\varphi)=F{'}\cos\varphi-F\sin\varphi,\\
    F{'}&=&\partial_\varphi F.
\end{eqnarray}

These basis vectors are related to the unit vectors by the scale factors,
\begin{eqnarray}
\label{Eq:ScaleFt}
h_{r}&=&[F^2\cos^2\varphi+
    \sin^2\varphi ]^{1/2} \nonumber \\
h_{\psi}&=&
    (\rho+rF\cos\varphi) \\
h_{\varphi}&=& r[
    \Im^2
    +\cos^2\varphi ]^{1/2}= rh \nonumber 
\end{eqnarray}
where, for analogy with the EC model, the metric tensor element $h_\varphi$ is renamed as $rh$.

The metric tensor must be built to relate the covariant (subscript) and contravariant (superscript) spaces. The elements of the metric tensor are,
\begin{eqnarray}
\label{Eq:MetricElm}
g_{rr}&=& \vec{\varepsilon}_{r} \cdot \vec{\varepsilon}_{r}= 
    [F^2\cos^2\varphi+
    \sin^2\varphi] ,  \\
g_{\psi\psi}&=&\vec{\varepsilon}_{\psi} \cdot \vec{\varepsilon}_{\psi} =
    [\rho+rF\cos\varphi]^2 = h_{\psi}^2,
    \\
g_{\varphi\varphi}&=&\vec{\varepsilon}_{\varphi} \cdot \vec{\varepsilon}_{\varphi} =
    r^2[\Im^2
    +\cos^2\varphi ]=r^2h^2,  \\
g_{r\psi}&=&g_{\psi r} = \vec{\varepsilon}_{r} \cdot \vec{\varepsilon}_{\psi}=
   0, \\
g_{r\varphi}&=&g_{\varphi r} = \vec{\varepsilon}_{r} \cdot \vec{\varepsilon}_{\varphi}= 
    r\cos\varphi[\Im F +
    \sin\varphi] = r \overline{g}_{r\varphi},      \\
\label{Eq:metricElmF}
g_{\psi\varphi}&=&g_{\psi \varphi} = 
    \vec{\varepsilon}_{\psi}\cdot \vec{\varepsilon}_{\varphi}=0.
\end{eqnarray}

Note we have renamed $g_{r\varphi}=r\overline{g}_{r\varphi}$ to separate the radial and the angular dependency. 

The metric is then,
\begin{equation}
g^{1/2} 
=[g_{rr}g_{\psi\psi}g_{\varphi\varphi}-g_{r\varphi}^2g_{\psi\psi}]^{1/2}=
    r(\rho+rF\cos\varphi)
    [F\cos^2\varphi-\Im\sin\varphi]= 
rh_\psi \overline{g} .
\label{Eq:metric}
\end{equation}
Note also that for the cylindrical approximation the $h_\psi =1 $ and the $\overline{g}=\delta$ (EC model) or $1$ (CC model).

The covariant components of the metric are:
\begin{eqnarray}
g^{rr} &=&\frac{g_{\psi\psi}g_{\varphi\varphi}}{g}
    = \frac{h^2}{\overline{g}}\\
g^{\psi\psi} &=& 
    \frac{g_{rr}g_{\varphi\varphi}-g_{r\varphi}^2}{g}
    =\frac{1}{h_\psi} \\
g^{\varphi\varphi} &=& \frac{g_{rr}g_{\psi\psi}}{g}
    =\frac{h}{r^2\overline{g}}\\
g^{r\varphi} &=& g^{\varphi r} 
    =-\frac{g_{r\varphi}g_{\psi\psi} }{g}
    =-\frac{\overline{g}_{r\varphi}}{r\overline{g}}
\end{eqnarray}

The final step is to define operators that allow us to solve the MHD equations and obtain the physical quantities that characterize heliospheric flux ropes (see NC18 for more details). The \textit{divergence} of the magnetic field is given by
\begin{equation}
\bigtriangledown\cdot \vec{B}=\frac{1}{g^{1/2}}\frac{\partial}{\partial q^{k}}(g^{1/2}B^{k}_{c}) 
\label{Eq:div}
\end{equation}
where $B^{k}_{c}$ are the non-scaled contravariant components of the magnetic field, and $q^{k}$= ($r,\psi,\varphi$). Same equation is used for the current density ($\bigtriangledown\cdot\overrightarrow{j}=0$).

The curl components for the magnetic field are given by
\begin{equation}
(\bigtriangledown\times \vec{B})^{i}=\frac{1}{g^{1/2}}\varepsilon^{ijk} \frac{\partial }{\partial q^{j}}(g_{kl}B^{l}_{c})
\label{Eq:curl}
\end{equation}
where $\varepsilon^{ijik}$ are the Levi-Civita Coefficients. 

Assuming ($0,B_{c}^{y}$,$B_{c}^{\varphi}$) and ($j_c^r$,$j_{c}^{y}$,$j_{c}^{\varphi}$) vector components of the magnetic field and current density, the equations to solve are obtained from Ampere's Law, and Gauss's law for magnetism for a stationary case and imposing the continuity equation for the currents. Based on the geometry, $B_c^r=0$ can be assumed and the equation system will be:
\begin{eqnarray}
\label{Eq:EQ1}
\partial_\varphi(g^{1/2}B_c^\varphi) &=& 0 ,\\
\label{Eq:EQ2}
\partial_\varphi (g_{\psi\psi}B_c^{\psi}) &=&   -g^{1/2}\mu_0 j_c^r ,\\
\label{Eq:EQ3}
\partial_\varphi(g_{r\varphi}B_c^{\varphi}) -
    \partial_r (g_{\varphi\varphi}B_c^{\varphi}) 
    &=&    g^{1/2}\mu_0 j_c^\psi ,\\
\label{Eq:EQ4}
\partial_r(g_{\psi\psi}B_c^{\psi})
    &=&    g^{1/2}\mu_0 j_c^\varphi ,   \\
\label{Eq:EQ5}
\partial_r(g^{1/2}j_c^r) + \partial_\varphi(g^{1/2}j_c^\varphi) &=& 0.
\end{eqnarray}
where the magnetic field and current density components are the non-scaled contravariant coefficients of $\vec\varepsilon_{i}$ and they should be scaled using the scale factors, equations~(\ref{Eq:ScaleFt}). 

Then, directly from equation~(\ref{Eq:EQ1}), the solution for  the poloidal magnetic field component must be,
\begin{equation}
    B^\varphi_c (r,\varphi) = \frac{\overline{B}_c^\varphi(r)}{h_\psi\overline{g}}
\end{equation}
where henceforth $\overline{B}_c^\varphi=\overline{B}_c^\varphi(r)$.

Now solving for $B_c^\varphi$, the equation~(\ref{Eq:EQ3}),
\begin{eqnarray}
   \partial_r (\overline{B}_c^{\varphi}) 
    +
    \bigg[\frac{1}{r}\chi(\varphi)-\frac{F\cos\varphi}{h_\psi}\sigma(\varphi)\bigg] \overline{B}_c^{\varphi}
    &=&  - \mu_0 \frac{h^2_\psi\overline{g}^2}{r h^2} j_c^\psi, 
\end{eqnarray}
with,
\begin{eqnarray}
    \chi(\varphi)&=& 2-\frac{\partial_\varphi \overline{g}_{r\varphi}}{h^2}+\frac{\overline{g}_{r\varphi}\partial_\varphi\overline{g}}{\overline{g}h^2}\\
    \sigma(\varphi) &=&1 +\frac{\overline{g}_{r\varphi}}{h^2}\frac{\Im}{F\cos\varphi}.
\end{eqnarray}

The solution to the equation is, 
\begin{equation}
\overline{B}_c^{\varphi}
    =   - \mu_0 \frac{h_\psi^{\sigma}}{r^{\chi}} \frac{ \overline{g}^2}
   { h^2 }\int_0^r \frac{r^{\chi-1}}{h_\psi^{\sigma-2}} j_c^\psi(r,\varphi) dr',
\end{equation}
where $j^\psi_c(r,\varphi)$ should be such to ensure that $\overline{B}_c^\varphi(r)$ does not depend on the poloidal coordinate. Thus, 
\begin{equation}
    j_c^\psi(r,\varphi)=\frac{h^2}{\overline{g}^2}\frac{h_\psi^{\sigma-2}}{r^{\chi-1}}\partial_r\big[\frac{r^{\chi}}{h_\psi^\sigma}k(r)\big]=
    \frac{h^2}
    {\overline{g}^2 h_\psi^2} \bigg[k(r)(\chi-\frac{r}{h_\psi}\sigma) +k{'}(r)r \bigg],
\end{equation}
where $k(r)$ is an arbitrary function solely dependent on $r$. The non-scaled poloidal magnetic field component is,
\begin{equation}
    B_c^\varphi = \frac{\overline{B}_c^\varphi}{h_\psi \overline{g}} = - \mu_0 \frac{k(r)}{h_\psi \overline{g}}
\end{equation}

Now solving, from equation~(\ref{Eq:EQ5}), the poloidal current density component, 
\begin{equation}
j_c^\varphi = 
\frac{\partial_r(rf(r))}{r\overline{g}}
    ,
\end{equation}
and the radial current density component is,
\begin{equation}
    j_c^r =-\frac{\Im}{h_\psi \overline{g}} f(r)
\end{equation}

Note that, in the particular case of $F(\varphi)=constant$ and cylindrical geometry, $j^r = 0$ would be a solution of the equations. This was the solution expressed in the case of the EC and CC models. 

Now, to solve for the toroidal magnetic field component,  we can impose  $B_c^\psi(r,\varphi)=\frac{\overline{B}_c^\psi(r)}{h_\psi}$, then the scaled toroidal magnetic field component,
\begin{equation}
    \overline{B}_c^{\psi}(r) = -\mu_0 f(r)\bigg|_0^r,
\end{equation}
where we will impose that the value of the central magnetic field will decrease with the radial distance to reach a value at $r=R$ that may be canceled or a scaled value of the central magnetic field. This boundary condition may be arbitrary set. We will rename this function as
\begin{equation}
    f_r(r) = f(r)\big|_0^r.
\end{equation} 

Then, the non-scaled toroidal magnetic field component is,
\begin{equation}
    B_c^\psi = -\mu_0 \frac{1}{h_\psi} f_r(r)\big|_0^r.
\end{equation}

The scaled magnetic field components are,
\begin{eqnarray}
    B^r(r,\varphi) &=&0 \nonumber \\
    B^\psi (r,\varphi) &=&  h_\psi B_c^{\psi}(r,\varphi)=-\mu_0  \bigg[f(r)\big|_0^r\bigg]\\
    B^\varphi(r,\varphi)  &=& rh B_c^{\varphi}(r,\varphi) 
     = - \mu_0 \frac{h}{h_\psi \overline{g}}rk(r). \nonumber
\label{Eq:Beq}
\end{eqnarray}

The scaled current density components are,
\begin{eqnarray}
    j^r(r,\varphi)  &=& -h_r j_c^r(r,\varphi) =  \frac{h_r \Im}{\overline{g}h_\psi}f(r)
     \nonumber \\
    j^\psi(r,\varphi) &=&h_\psi j_c^\psi(r,\varphi)=\frac{h^2}
    {\overline{g}^2 h_\psi} \bigg[k(r)(\chi-\frac{r}{h_\psi}\sigma) +k{'}(r)r \bigg],  \\
    j^\varphi(r,\varphi) &=& rh j_c^\varphi(r,\varphi) =   h\frac{\partial_r(rf(r))}
    {\overline{g}}. \nonumber
\label{Eq:Jeq}
\end{eqnarray}
The set of equations (\ref{Eq:Beq}) and (\ref{Eq:Jeq}) are the general solution of the magnetic field for the radial profile of the current density components. Table~\ref{tab:list} summarizes the geometrical factors and parameterized equations that directly impact in the above model equations. Thus, for any geometry consistent with the coordinate system, Equation~(\ref{Eq:CoordSyst}), and a chosen radial profile of the current density component, the solution can be found.  

\begin{table}[]
    \centering 
    \begin{tabular}{|l|l|}
    \hline
   & GEOMETRICAL FACTORS \\
    & $\&$ 
    PARAMETRIC FUNCT. 
        \\
    \hline \hline    
     1.& $\Im(\varphi)=F{'}(\varphi)\cos\varphi -F(\varphi)\sin\varphi$ \\
    2. &$h_r=[F^2\cos^2\varphi+\sin^2\varphi]^{1/2}$ 
        \\
    3. &$h=[\Im^2(\varphi)+\cos^2\varphi]^{1/2}$ 
        \\
    4.& $h_\psi = (\rho+rFcos\varphi)$ 
        \\
    5.& $\overline{g} = [F\cos^2\varphi-\Im\sin\varphi]$ 
        \\
    6. & $\overline{g}_{r\varphi} = \cos\varphi[\Im F +    \sin\varphi] $ 
        \\
    \hline
    \end{tabular}
    \caption{Geometrical factors, including scale factors and metric, and parameterized equations needed to build the model equations.}
    \label{tab:list}
\end{table}

Similar to previous papers, the relevant physical quantities to study these structures in Heliophysics can be obtained from the previous equations. Below we list some of them, such as the magnetic fluxes,
\begin{eqnarray}
\label{Eq:Flujo_psi}
\Phi_\psi &=& \int B_c^{\psi}dA_{\psi}= \int B_c^{\psi}g^{1/2}drd\varphi  
= \mu_0 \int_0^\varphi \overline{g} d\varphi \int_0^R rf(r) \bigg|_0^r dr
\\
\label{Eq:Flujo_varphi}
\Phi_\varphi &=& \int             B_c^{\varphi}dA_{\varphi}    =\int B_c^{\varphi} g^{1/2}drd\psi =
2\pi \mu_0 \int_0^R rk(r) dr .
    \label{Eq:Flujo_pol}   
\end{eqnarray}
Obviously, the toroidal magnetic field will depend on the cross-section geometry, while the poloidal magnetic flux is not altered.  

The magnetic energy,
\begin{equation}
    W = \int \frac{B^2}{2\mu_0}dV 
    =\int \frac{B^2}{2\mu_0} g^{1/2} drd\psi d\varphi
\end{equation}

For completeness we include the magnetic helicity, calculated from the dot product with the magnetic potential \citep[see][]{Woltjer1958,Taylor1974, Brown1999,Arfken2005},
\begin{equation}
    H = \int \vec{B}\cdot \vec{A} ~ dV = \int g^{1/2} g_{ik} B_c^i A_c^k  ~drd\psi d\varphi
\end{equation}
The vector potential also requires an additional calculation that will depend on the distortion and may also require numerical solutions. 

Finally, the non-scaled covariant cross-product that will lead to the components of the Lorentz force are,
\begin{equation}
\label{Eq:DQ2}
(\vec{j} \times \vec{B})|_{i,c}=g^{1/2}\epsilon_{ijk}j^{j}_{c}B^{k}_{c}. \\
\end{equation}

Expanding the equation, now there are toroidal and poloidal components in the internal Lorentz forces due to the distortion and curvature:
\begin{eqnarray}
    (\vec{j}\times\vec{B})|_{c,r}&=&rh_\psi \overline{g}(j_c^\psi B_c^\varphi - j_c^\varphi B_c^\psi ) 
    \\
    (\vec{j}\times\vec{B})|_{c,\psi}&=& - rh_\psi \overline{g}(j_c^r B_c^\varphi)
    \\
    (\vec{j}\times\vec{B})|_{c,\varphi}&=& rh_\psi \overline{g}(j_c^r B_c^\psi)  
\end{eqnarray}

This model establishes the mathematical formulation to explore distortions in the observed 3D flux ropes, 
but the solution of the above equations system would change depending on the kind of distortion of the flux-rope under study as well as the internal distribution of the current densities.  
Following the approach in \citet[][]{Nieves-Chinchilla2016,Nieves-Chinchilla2018ApJ}, in the next subsection, we evaluate the solutions to the equations based on the general radial variation of the current densities with the specific specific geometry based on the $F=\delta(1 - \lambda \cos\varphi)$ for a highly curved and cylindrical flux rope. 

\subsection{A general case of radial variation of the current density components}
In agreement with previous papers, the current density components could be selected using the radial polynomial function with arbitrary coefficients. Here, we are going to simplify the problem and select one series term to develop the problem. Thus,
\begin{eqnarray}
    k(r)&=&\beta_m r^{m} \text{, with m}\ge 0,\\
\label{Eq:kr}
    f(r)&=&-\alpha_n r^{n+1} \text{, with n}\ge 1,
\label{Eq:fr}
\end{eqnarray}
where $\alpha_n$ and $\beta_m$ are the two coefficients parameters of the model, and the $m,n$ indexes determine the radial profile of the current density components and eventually the magnetic field.

The scaled current density components are,
\begin{eqnarray}
    j^r(r,\varphi)  &=&  h_r \frac{\Im}{h_\psi \overline{g}}\alpha_n r^{n+1}, \nonumber \\
    j^\psi(r,\varphi) &=& \frac{h^2h_\psi}{\overline{g}^2}\beta_m r^m[\chi+m-\frac{r}{h_\psi}\sigma],  \\
    j^\varphi(r,\varphi) &=& -\frac{h}{ \overline{g}}(n+1)\alpha_n r^{n}. \nonumber
\label{Eq:modelDTcurrents}
\end{eqnarray}

The scaled magnetic field components are,
\begin{eqnarray}
    B^r(r,\varphi) &=&0 \nonumber \\
    B^\psi (r,\varphi) &=& \mu_0 \alpha_n  [\tau R^{n+1} -  r^{n+1}]\\
    B^\varphi(r,\varphi)  &=&  - \frac{h}{h_\psi \overline{g}} \beta_m r^{m+1}. \nonumber
\label{Eq:modelDTfields}    
\end{eqnarray}

The above equations could be simplified to one single term of the polynomial series and parameterized to the parameters, $C_{nm}=\frac{\alpha_n}{\beta_m}R^{n-m}$ and $B_n = \mu_0 \frac{\alpha_n}{R^{n+1}}$. Thus, the magnetic field components,
\begin{eqnarray}
    B^\psi (r,\varphi) &=& B_n  [\tau -  \overline{r}^{n+1}]\\
    B^\varphi(r,\varphi)  &=&  - \frac{h}{h_\psi \overline{g}}\frac{B_n}{C_{nm}} \overline{r}^{m+1}, \nonumber
\end{eqnarray}
and, the current density components are, 
\begin{eqnarray}
    j^r &=& \frac{h_r}{h_\psi}\frac{\Im}{F}\alpha_n R^n \overline{r}^{n+1},
    \nonumber\\
    j^\psi &=& \frac{h^2h_\psi}{\overline{g}}\frac{\alpha_n}{C_{nm}}R^n \overline{r}^m[\chi+m-\overline{r}\frac{R}{h_\psi}\sigma],
    \\
    j^\varphi&=& -\frac{h}{h_\psi \overline{g}}\alpha_n R^n \overline{r}^{n}[F\cos\varphi R \overline{r}+(n+1)h_\psi], \nonumber
\end{eqnarray}
where $\overline{r}=r/R$.
   
\begin{table}[]
    \centering
    \small
    \begin{tabular}{|l|l|l|}
       \hline \hline
       CIRCULAR & ELLIPTIC & DISTORTED
       \\ 
       F= 1 & 
       F=$\delta$ & 
       F=$\delta(1-\lambda\cos\varphi)$ 
       \\
       \hline
        &&\\
       $\Im= -\sin\varphi$ & $\Im=-\delta\sin\varphi$ & $\Im=\delta \sin \varphi[2\lambda\cos\varphi-1]$
       \\
       $h_r=1$ & $h_r=[\sin^2\varphi+\delta^2\cos^2\varphi]^{1/2}$ & $h_r=[\sin^2\varphi+F^2 \cos^2\varphi]^{1/2}$ 
        \\
        $h_\psi=(\rho+r\cos\varphi)$ &  $h_\psi=(\rho+r\delta\cos\varphi)$ &
        $h_\psi=(\rho+rF\cos\varphi)$ 
        \\
        $h=1$ &
        $h=[\delta^2\sin^2\varphi +\cos^2\varphi]^{1/2}$ & $h=[F^2\sin^2\varphi +\cos^2\varphi]^{1/2}$ 
        \\
        $\overline{g}=1$ &
        $\overline{g}=\delta$ &
        $\overline{g}= F-\delta\lambda\sin^2\varphi\cos\varphi$ 
        \\
        & & \\
        \hline
        & &    \\
        $B^r = 0$ &
        $B^r = 0$ &
        $B^r = 0$ 
        \\
        $B^\psi = B_n  [\tau -  \overline{r}^{n+1}]$ &
        $B^\psi = B_n  [\tau -  \overline{r}^{n+1}]$ &
        $B^\psi = B_n  [\tau -  \overline{r}^{n+1}]$ 
        \\ 
        $B^\varphi= -  \frac{1}{h_\psi}\frac{B_n}{C_{nm}} \overline{r}^{m+1}$ &
        $B^\varphi =-  \frac{h}{\delta h_\psi}\frac{B_n}{C_{nm}} \overline{r}^{m+1}$ &
        $B^\varphi = - \frac{h}{ \overline{g} h_\psi}\frac{B_n}{C_{nm}} \overline{r}^{m+1}$ 
        \\ 
        & & \\
        &&\\
\hline
\hline
    \end{tabular}
    \caption{Functions, scale factors, parametric functions and model equations associated to the geometry associated to $F=1$, $F=\delta$, and $F=\delta(1\pm\lambda\cos\varphi)$.} 
    \label{tab:Table2}
\end{table}

\begin{figure}
    \centering
   \includegraphics[width=0.85\textwidth]{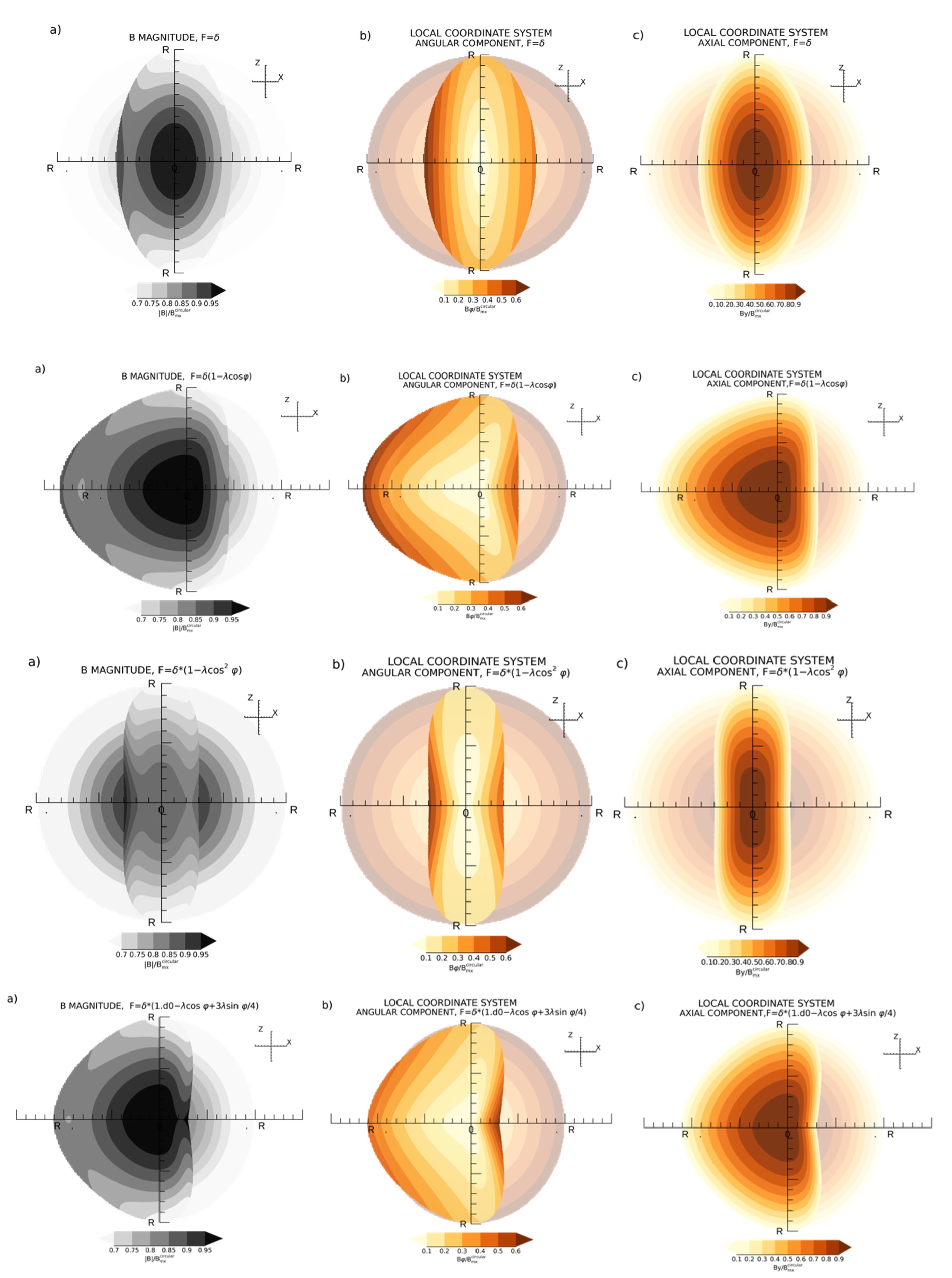}
    \caption{Each column in this array of plots shows the magnetic field magnitude distribution (first column), and poloidal (second column) and axial (third column) components for the four $F-$functions displayed in Figure~\ref{fig:figure2}. From the top (I) F=$\delta$ with $\delta=0.5$ and $\rho=1.5R$; (II) $F=\delta(1-\lambda cos\varphi)$ with $\delta=0.9$, $\lambda$=0.4 and $\rho=2R$; (III) $F=\delta(1-\lambda\cos^2\varphi)$ with $\delta=0.5$,  $\lambda$=0.4  and $\rho=2R$; and (IV) $\delta(1-\lambda\cos\varphi+3\lambda\sin\varphi/4)$ with $\delta=0.9$ and $\lambda$=0.4. In all simulations, $\rho=3R$. Each case is displayed on top of the corresponding quantity for the semitransparent circular cross-section distribution to highlight the change due to the distortion. The torus center is located at the left of each plot.   }
    \label{fig:figure3}
\end{figure}

Table~\ref{tab:Table2} includes the analytical solution for the magnetic field and current density components for the circular, elliptical and the distorted ($F=\delta(1\pm\lambda\cos\varphi)$) case. The table also includes all geometrical factors and solutions for the parameterized equations listed in Table~\ref{tab:list}. The same exercise can be done for the distortions shown in Figure~\ref{fig:figure2}. Thus, the plot array in Figure~\ref{fig:figure3} shows the total magnetic field distribution (first column), poloidal (second column) and axial (third column) magnetic field components distribution for the different four distorted cross-sections in the same order of Figure~\ref{fig:figure2}. Figure~\ref{fig:figure2}(a) displays the case of F=$\delta$; (b) $F=\delta(1-\lambda cos\varphi)$, (c) $F=\delta(1-\lambda\cos^2\varphi)$, and (d) $\delta(1-\lambda\cos\varphi+3\lambda\sin\varphi/4)$. Each case is displayed on top of the corresponding quantity for the semitransparent circular cross-section distribution to highlight the change due to the distortion.  
In the simulation, all cases we have considered use $\rho=3R$, $\lambda=0.4$ and $\delta = 0.5$ excepting of Figure~\ref{fig:figure2}II that considers $\delta=0.9$. 

For the magnetic field strength, in all cases the contour lines indicate the maximum magnetic field strength that remains at the $\rho$ distance to the torus center. The torus center is located at the left of each plot, thus due to the curvature it is observed an increase in the magnetic field strength for all cases. In the symmetric cases, (a) and (c) is more significant. In the case of a spacecraft crossing the structure, the magnetic field configuration would display an asymmetric magnetic field magnitude profile with an increase at the rear part of the structure. This scenario discussed in \citet{Nieves-Chinchilla2018SolPhy}, Figure 11, would describe up to 22$\%$ of the ICMEs observed by the Wind spacecraft in the studied interval. The specific case of the event observed in November 6, 2000 (doy 311), Figure 6c in the paper, depicts a back compression in the magnetic field strength in spite of the structure showing an expansion velocity $V_{exp}=$58 $km~s^{-1}$. In the case of the asymmetric cross-section distortions, cases (b) and (c) in the Figure~\ref{fig:figure3}, there is a bilateral compression. Same effect due to the curvature plus the front compression due to the distortion. The magnetic field profile may result in a symmetric magnetic field strength and it would be difficult to decipher signatures of distortion.   

For the magnetic field components, there is an increase in the magnetic field strength due to the axis curvature. Just for the asymmetric cross-section distortions, there is an increase in the poloidal and axial magnetic field at the front and rear parts of the structure. 
The next section will show more insights for the specific case of $F=\delta(1\pm\lambda\cos\varphi)$ in the highly curved and cylindrical flux rope structure.

\section{Magnetic field imprints of the curved-distorted flux-rope}
As part of this paper, we have carried out the study of the implications of the distortion in the magnetic field configuration as observed by spacecraft crossing a flux-rope. The goal is to identify what in situ signatures could provide insights of distortion and curvature.  
We think that it will be premature and artificial to fit the model to the data since we could increase the number of parameters to improve the goodness parameters. However, we think htat the first apporach would be to learn from the model to identify signatures of distortion. This exercise would be similar but opposite to the study carried out for \citet{Nieves-Chinchilla2018SolPhy} that looked for signatures in the ICMEs observed by Wind for 20 years to identify the distortion, expansion or curvature signatures. In our case, we will simulate the trajectory of a spacecraft to train our brain and learn how to identify signatures of distortion or curvature in the real data. This would be the preliminary exercise to eventually train a machine. 

Following in the section, we will first define the trajectory of the spacecraft taking into consideration the geometry of the distortion, so as to generate synthetic magnetic field profiles. Then, we will discuss the implications of the curvature and distortion in the in-situ observations of the magnetic field as observed by a satellite. For the experiment we will use one of the functions we discussed in the previous section $F=\delta(1-\lambda \cos \varphi)$. We will cross two different curvatures and we will map the magnetic field magnitude and components to evaluate the deviations from the expected in-situ signatures that a magnetometer would record if crossing a cylinder with circular crossection geometry. 
We then will look for such signatures in the real data. This exercise will pave the way to develop a more sophisticate model to identify such signatures. 

\subsection{Spacecraft trajectory}
This section is dedicated to evaluate the effect of the distortion in the 3D reconstructions based on in-situ data. The trajectory of the spacecraft is defined by the location of the spacecraft at the entrance of the flux rope as indicated by \citet{Nieves-Chinchilla2018ApJ},
\begin{equation}
\label{Eq:x0R}
    x_0 = \frac{v_{sw}(t_{t}-t_0)}{2} - \frac{F_1}{F_2}z_0,
\end{equation}
In the case of simulated spacecraft trajectory, the transit time will be,
\begin{equation}
\label{Eq:ts}
t_{s}=\frac{2F}{v_{sw}} \frac{\sqrt{R^2F_2  -y_0^2F_3^2}}{F_2},
\end{equation}
where,
\begin{eqnarray}
    F_1&=&(F^2-1)\cos\phi\cos\theta\sin\xi\cos\xi + 
    a\sin\phi\sin\theta\cos\theta \nonumber\\
    F_2&=&b\cos^2\phi 
    +a\sin^2\phi\sin^2\theta+
    2(F^2-1)\cos\phi\sin\phi\sin\theta\sin\xi\cos\xi,
    \nonumber \\
    F_3&=&\sqrt{1-\sin^2\phi\cos^2\theta} \nonumber
\end{eqnarray}
and,
\begin{eqnarray}
    a=[F^2\cos^2\xi+\sin^2\xi] \nonumber \\
    b=[F^2\sin^2\xi + \cos^2\xi]. 
\end{eqnarray}
Note that we have corrected here an error in equation~(51) in  \citet{Nieves-Chinchilla2018ApJ}. Note that in the case of a reconstruction the spacecraft transit time, $t_s$, and the bulk speed, $v_{sw}$, are obtained from the observations and the flux-rope radius, $R$, is obtained as a deduced output parameter. In the case of synthetic data, the $R$ and $v_{sw}$ are input parameters, and $t_s$ is the output parameter. 

\subsection{Implications in magnetic field imprints at a spacecraft crossing a distorted flux rope with different curvatures and impact parameter}
\label{sec:section3}
In this section, we would like to understand the implications to analyze the resultant magnetic field configuration when the structure is crossed by a spacecraft with \textbf{the flux rope having?} different curvature of the axis. 
For this first experiment, we have selected the $m,n$-pair [0,1] and $F=\delta(1-\lambda\cos\varphi)$ to describe the cross-section geometry. The magnetic field model equations are,
\begin{eqnarray}
B^\psi(r,\varphi) &=& B_1[\tau - 
    \overline{r}^{2}], \nonumber\\
B^\varphi(r,\varphi) &=&
    - \frac{h}{h_{\psi}\overline{g}}
    \frac{B_1}{C_{10}} \overline{r}. 
\end{eqnarray}
with,
\begin{eqnarray}
    \Im&=&\delta \sin \varphi[1+2\lambda\cos\varphi], \nonumber\\
    h&=&[\delta(1-\lambda\cos\varphi]^2\sin^2\varphi + \cos^2\varphi]^{1/2} ,\nonumber \\
    h_\psi&=&(\rho+r\delta(1-\lambda\cos\varphi)\cos\varphi),\\
    \overline{g}&=& \delta(1-\lambda\cos\varphi), \nonumber
\end{eqnarray}
from table~\ref{tab:Table2}. The $B_1,\ C_{10}$, and $\tau$ are model parameters and $\overline{r}$ is the normalized cross-section radial distance. 

\begin{figure}
\centering
   \includegraphics[width=0.49\textwidth]{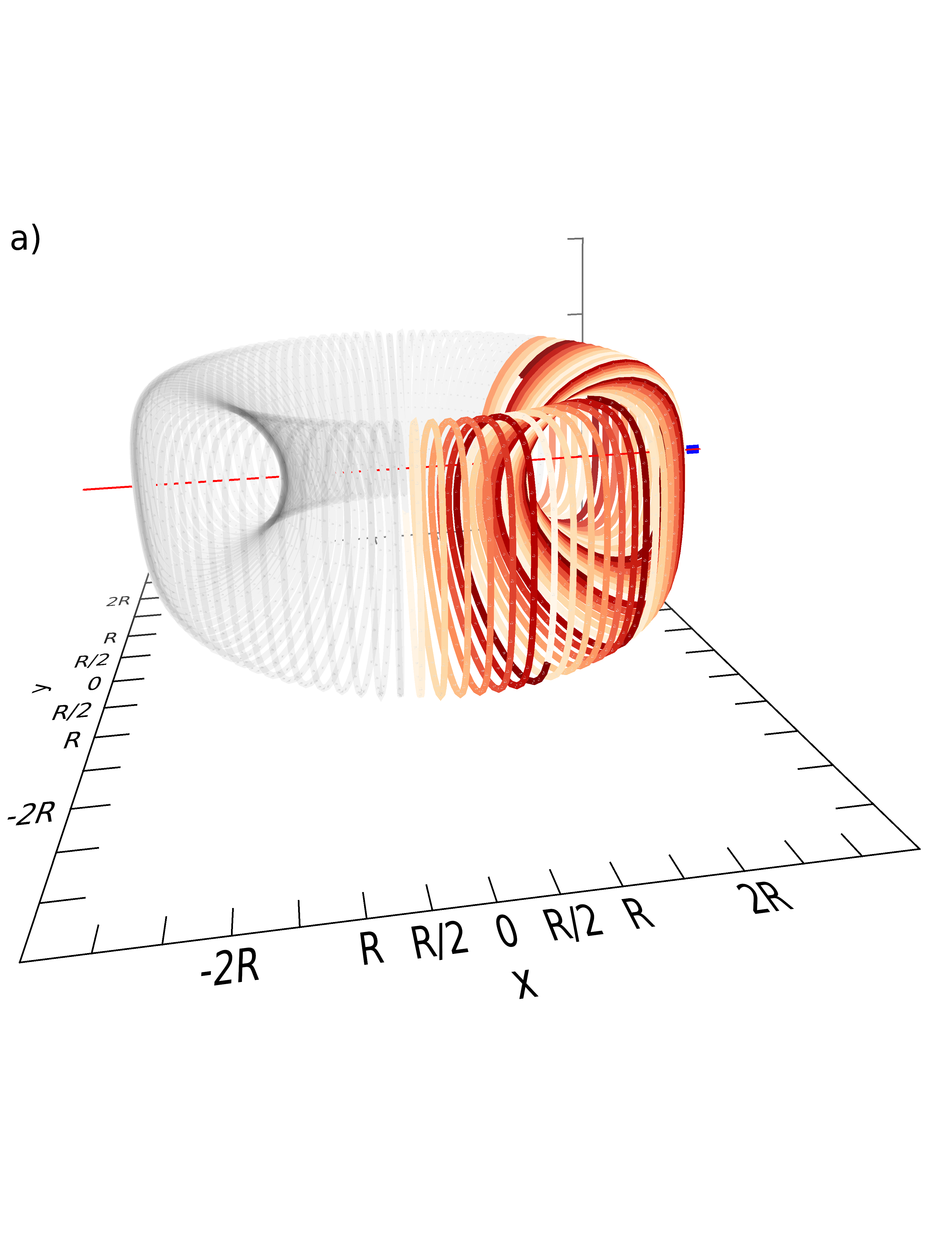}
    \includegraphics[width=0.49\textwidth]{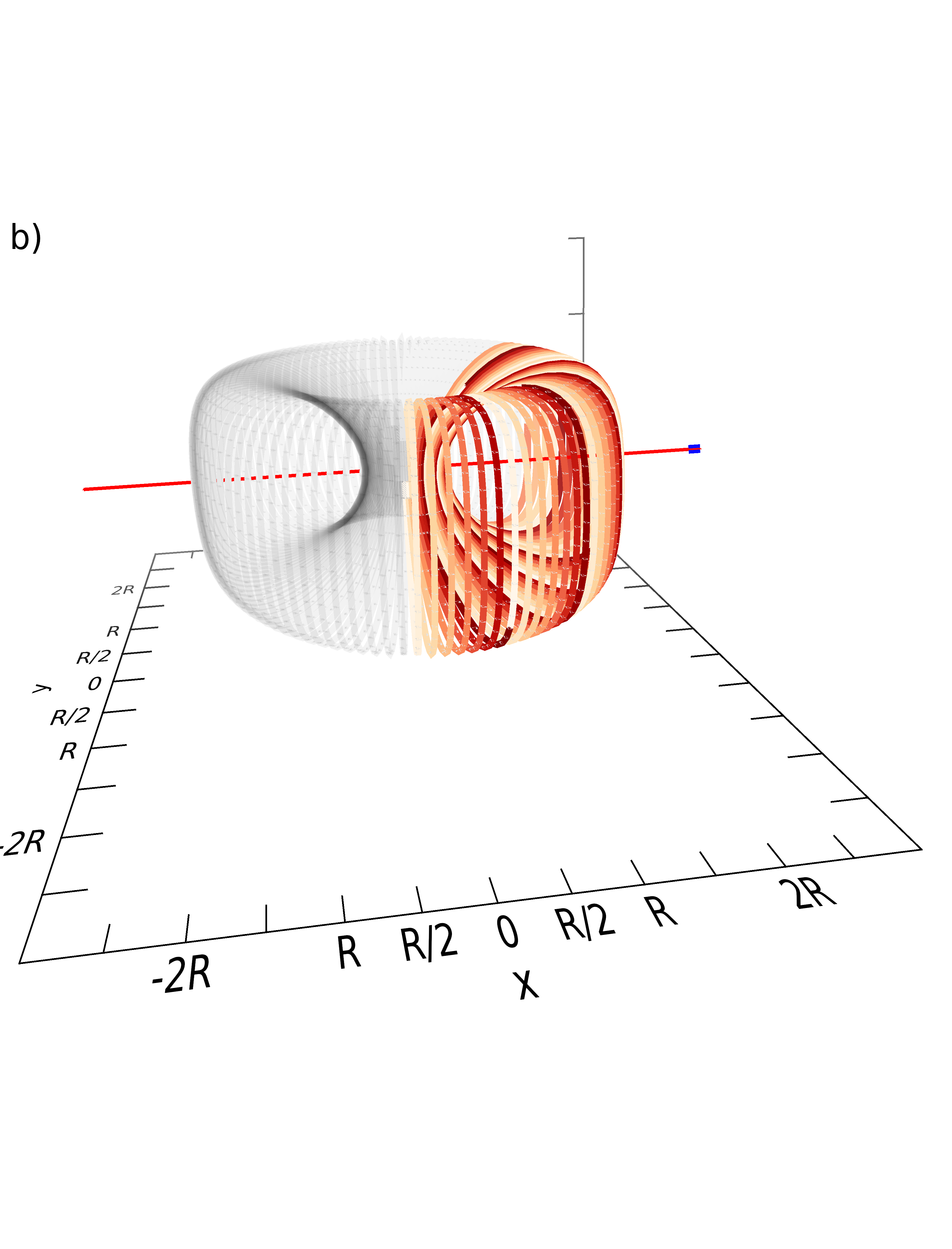}
    \caption{3D view of the toroidal flux-rope based on the geometry $F=\delta(1-\lambda cos\varphi)$ for (a) $\rho=1.5R$, and (b)$\rho=2.5R$. In both cases, $\delta=0.8$, $\lambda=0.4$. The red line indicates the trajectory of a simulated spacecraft. The blue dots indicate the spacecraft entrance \textbf{location?}. 
    }
    \label{fig:figure4}
\end{figure}

We have considered two toroidal geometries to compare with the cylindrical case. Figure~\ref{fig:figure4} illustrates the flux ropes with a central radius of $\rho=2.5R$ (Figure~\ref{fig:figure4}a) and $\rho=1.5R$ (Figure~\ref{fig:figure4}b) respectively, with R being the major radius of the cross-section. The tori have been partially colored at the front, simulating the front of a CME and indicating the trajectory of the spacecraft in red color with the spacecraft in blue color. The simulated spacecraft is crossing through the compressed front of the flux rope and is crossing through its center ($y_0=0$). The parameters for the simulation are:  $\delta=$0.8, $\lambda=$0.4, $\tau=$1.5, $C_{10}$=-1.5 (left-handed). 

\begin{figure}
\centering
   \includegraphics[width=\textwidth]{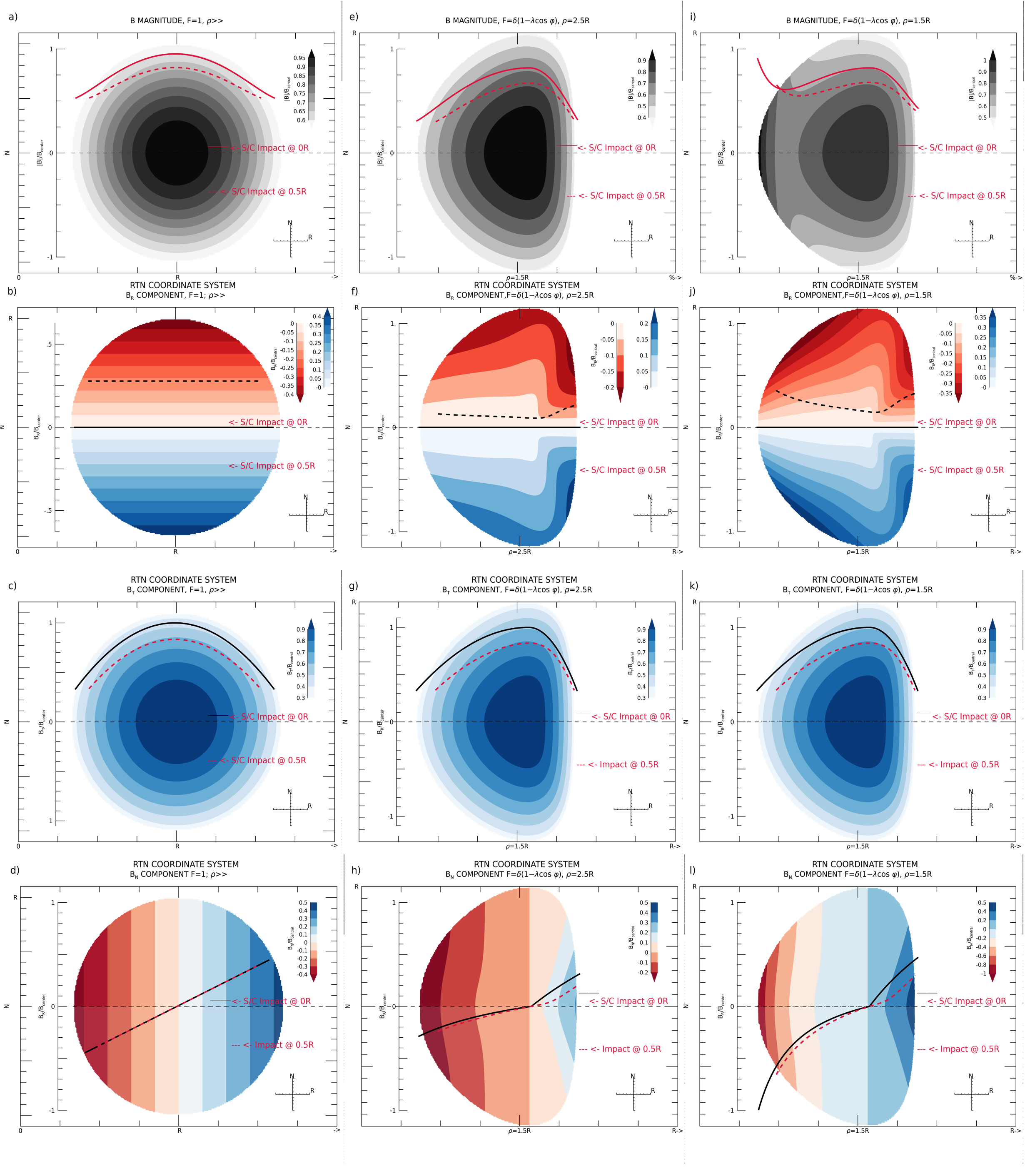}
   \caption{Contour plots the magnetic field strength (top row) and RTN coordinates (bottom three rows) distributed on the flux rope cross section in the cases of: (a-d) circular-cylindrical cross section; (e-h) distorted cross section with geometry $F=\delta(1-\lambda\cos\varphi)$ with large $\rho=2.5R$; and, (i-l) same geometrical distortion with $\rho=1.5R$ to enhance the effect of the curvature.  Overplot in each contour plot is shown the magnetic configuration observed by a spacecraft crossing through the flux rope center (colored black or red line) and at $50\%$ radial distance of the center, $y_0$=0.5R (colored black or red dashed line). 
   }
\label{fig:figure5}
\end{figure}

Thus, to evaluate the effect of the distortion as well as the curvature in the in-situ observations of the spacecraft crossing the structure, we have rotated the magnetic field components from the local coordinate system to, in this case, the RTN coordinate system. 
Figure~\ref{fig:figure5} depicts an array of plots for the two scenarios of Figure~\ref{fig:figure4} plus the case of the CC geometry that will serve as reference scenario to compare with. The figure shows the cross-section distribution of the magnetic field strength and RTN components for three different geometries: first column, neither distortion nor curvature (Figure~\ref{fig:figure5}(a)-(d), circular-cylindrical); second column, cross-section distorted but not curved, (Figure~\ref{fig:figure5}(e)-(f), $F=\delta(1-\lambda\cos\varphi)$ with $\rho=2.5R$); and, the third column, same cross-section distortion and highly curved (Figure~\ref{fig:figure5}(i)-(l), $F=\delta(1-\lambda\cos\varphi)$ with $\rho=1.5R$. For each geometry, we have simulated the crossing of a spacecraft from the left side through the center, as illustrated in Figure~\ref{fig:figure4}, and over plotted with colored line (black or red) the expected configuration of the magnetic field strength or components, depending on the case. From the top it is shown the magnetic field strength, R-component, T-component and N-component at the bottom. To evaluate the effect of the spacecraft impact distance with respect to the center on the quantities, we have also simulated the crossing of a spacecraft at $y_0=0.5R$, which $50\%$ away from the center (colored dashed line). 

Comparing the magnetic field strength (top row Figure~\ref{fig:figure5}(a),(e), and (f)), we can observe that the distortion implies an increase of the asymmetry in the contour plot with a compression of the magnetic field at the front of the structure. The curvature also implies an increase of the magnetic field strength. This assessment is confirmed if we analyze the magnetic field configuration as observed by the simulated spacecraft magnetometer, red lines over plotted on the contour black plots. While in the circular-cylindrical case the magnetic configuration is symmetric, in the distorted cases the maximum is displaced toward the front of the flux-rope. In the case of the highly curved flux rope, there is a marked increase in the strength as the spacecraft is leaving the structure, back side Figure~\ref{fig:figure5}(i). Note that as we increase the distance of the spacecraft to the center, the maximum of the strength decreases as expected, but it is important to highlight that also the effect of the curvature is less relevant. 

Given the relative orientation of the flux rope axis to the spacecraft trajectory, the $B_T$ component has a very similar profile to the magnetic strength, Figure~\ref{fig:figure5}(c), (g), and (k). Due to the distortion, there is a displacement of the maximum toward the distorted area, as it is also observed in the magnetic field configuration (black lines). However, this component does not display any feature associated with the curvature.  

The case of the $B_R$ component is very particular. This component is the one that provides information about the spacecraft impact distance to the center. In general, as described largely in the literature \citep[see for instance,][]{Demoulin2013, Nieves-Chinchilla2018SolPhy}, this component is completely flat or curved with maximum in the center, depending on the orientation when the circular axial symmetry is assumed. The contour plot in Figure\ref{fig:figure5}(b) illustrates this case with constant surfaces perpendicular to cross section. Thus in this case, for the two spacecraft crossings simulated, we find the constant values for the spacecraft impact at the center, with $B_R = 0 nT$ and $B_R=constant$ in the case of $y_0=0.5R$. The chirality determines the sign, so in our left-handed flux rope, the bottom half provides positive $B_R$ values. In the case of crossing through the top half, the $B_R$ values would be negative. As we add distortion, Figure\ref{fig:figure5}(f), or curvature, Figure\ref{fig:figure5}(j), the contours indicate a sudden change shortly after the compressed front of the cross-section. This effect is illustrated with the crossing at $y_0=0.5R$ (black dashed line). In the case of distortion only, Figure\ref{fig:figure5}(f), $B_R$ remains almost constant right after the front compression. However, in the case of adding curvature, Figure\ref{fig:figure5}(j), it implies that in addition to the sudden change close to the compressed area, there is also an increase (in magnitude) at the back of the structure (black dashed line). Interestingly, both effects of curvature or distortion, go unnoticed in the case of the spacecraft crossing through the center (black solid line) and are more noticeable as the spacecraft impact distance increases (black dash line).

The $B_N$ component changes the polarity from north to south, Figure~\ref{fig:figure5}(d), (h), and (l). For the CC geometry, the change on the polarity occurs in the center of the cross-section but the distortion implies a displacement towards the front of the cross-section. Thus, for this specific case, the duration of the positive polarity of the $B_N$ component is almost half of the negative polarity. The effect of the curvature implies a rapid change as the spacecraft approaches the curved area. For the flux-rope orientation, the B{$_N$} component is the one that mostly contributes to the increase in the back side of the magnetic field strength. In this case, the increase of the impact distance (red dashed lines) to the center implies again a mitigation of the sudden change in the polarity that it is observed with the distortion and curvature.

\section{Brief Discussion about the in situ and remote-sensing observations of distorted structures and the implications on Space Weather forecasting}
\label{Sec:section4}

\begin{figure}
\centering
    \includegraphics[width=0.7\textwidth]{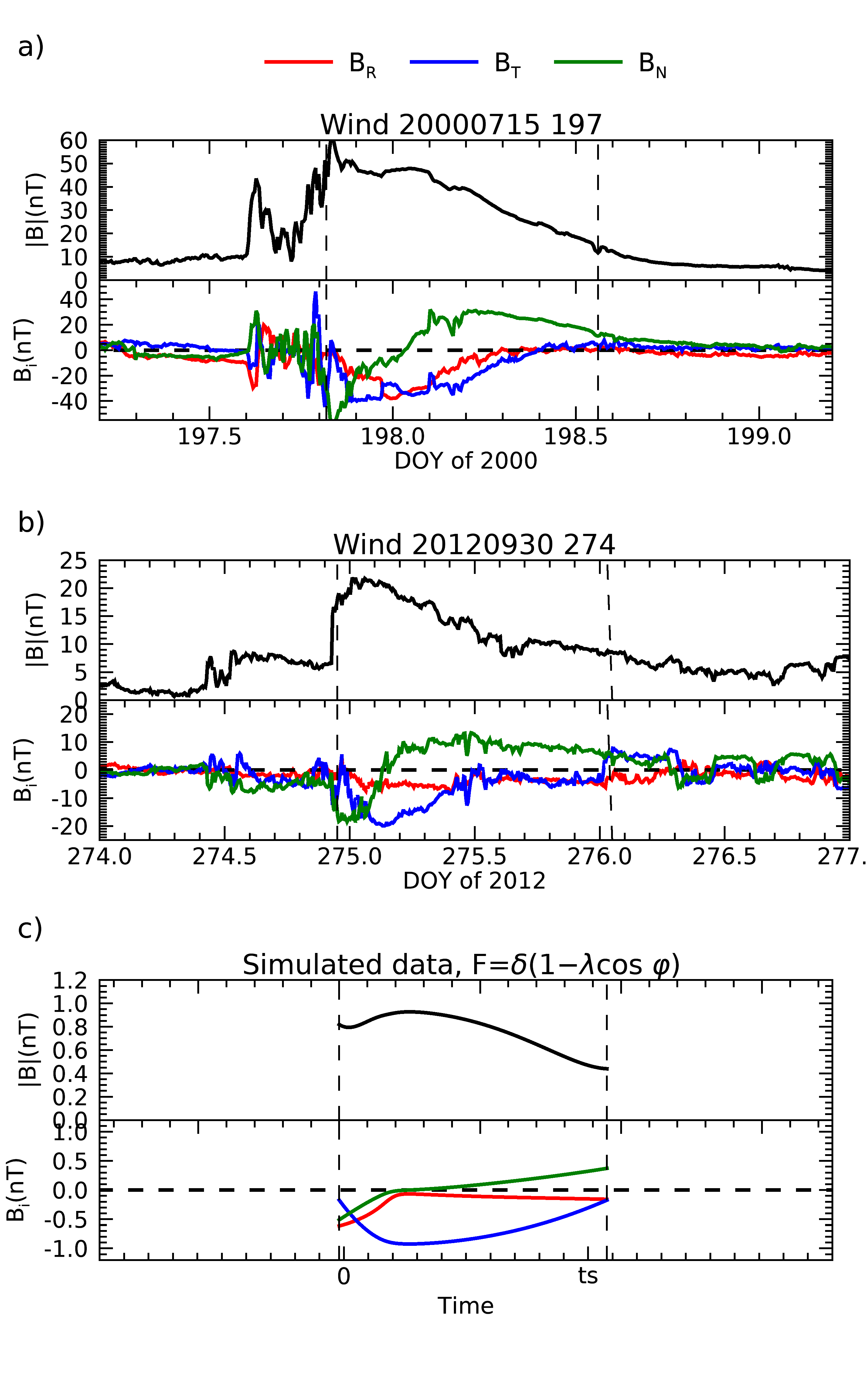} \\
   \caption{Interplanetary coronal mass ejections observed by the Wind spacecraft and simulated data. a) The Bastille Day event observed on July 15, 2000;, and, b) Event observed on September 30, 2012. The plots display the magnetic field magnitude and components in the RTN coordinate system. c) Simulated magnetic field observations of a spacecraft crossing a distorted flux rope. The selected function can be found at the top of the plot and the parameters values are $\rho=2.5R$, $\tau=1.2$, chirality = -1, $C_{10}=1.5$, $\lambda = 0.5$, and $\delta=0.5$. The two vertical dashed lines indicate the magnetic obstacle boundaries.}
   \label{fig:figure6}
\end{figure}

Figure~\ref{fig:figure6} displays two ICMEs observed by the Wind spacecraft \citep[see ICME Wind list, wind.gsfc.nasa.gov][]{Nieves-Chinchilla2018SolPhy}. These are two examples, where magnetic field signatures associated with distortions can be seen in in situ observations. At the top, the very well-known Bastille Day event observed on July 15, 2000, and, at the bottom, the event observed on September 30, 2012. The plots display the magnetic field magnitude at the top and the magnetic field components in RTN coordinate system at the bottom. The two vertical dashed lines indicate the magnetic obstacle boundaries. Based on the boundaries selected, both events display signatures associated with a flux rope, i.e. rotation of the magnetic field direction, a coherent magnetic structure and, not shown here, but associated with plasma signatures. 
In both cases, there is a clear compression in the magnetic field strength at the front of the structure. In the case of the Bastille day event, the $B_R$-component displays a clear change in the profile. The first part is curved to negative values and the second half rapidly reaches the zero flat value as could be described by Figure~\ref{fig:figure5}(f), in what would be a spacecraft crossing above but close to the flux-rope axis in a structure highly compressed at the front. Both events display asymmetric profiles similar to the magnitude in the $B_T$-component, which indicates that the structure axis is perpendicular to the spacecraft trajectory. Finally, the $B_N$-component, in both cases, crosses the polarity from south to north very close to the front, while the positive polarity phase of the $B_N$-component lasts almost twice in the case of the Bastille Day and almost three more times in the case of September 30, 2012 event. 

Figure~\ref{fig:figure6}(c) displays the simulated observations of a spacecraft crossing from the front, with perpendicular axis, to a distorted cross-section based on the function $F=\delta(1-\lambda \cos\varphi)$. The parameters selected are $\rho=2.5R$, $\tau=1.2$, chirality = -1, $C_{10}=1.5$, $\lambda = 0.5$, and $\delta=0.5$. Although the magnetic field components are magnitude are not identical to the real events, the visual comparative analysis exemplify the how this model may aid in the interpretation of actual observations exhibiting distortions. This exercise, based on visual inspection, accompanied by exploration of the physical problem using machine learning techniques  \citep[see for instance, ][]{dosSantos2020,Narock2022} will allow to make more accurate 3D reconstructions of the flux rope morphology, geometry, and physical parameters based on in situ observations.  


\begin{figure}
\centering
\setlength{\unitlength}{1cm}
    \begin{picture}(12.,8.39)   
        \put(0,0){\includegraphics[width=12cm,trim=2.8cm 2.2cm 3.9cm 2.1cm, clip]{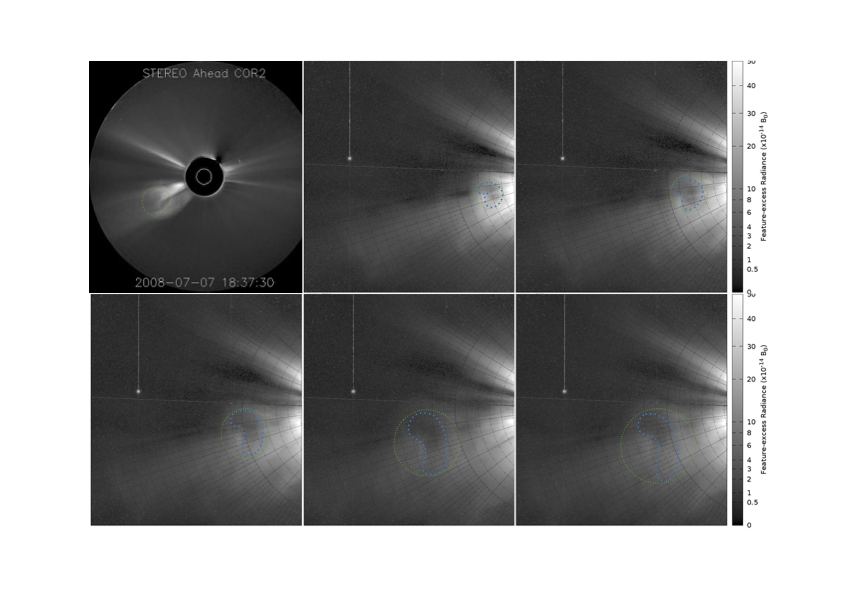}}
        \put(0.96,8.36){\textcolor{white}{\scriptsize \sffamily STEREO Ahead COR2}}
        \put(1.08,4.54){\textcolor{white}{\scriptsize \sffamily 2008\,-\,07\,-\,07~~18:37:30}}
        \put(4.9,8.36){\textcolor{white}{\scriptsize \sffamily STEREO Ahead HI1}}
        \put(4.9,4.54){\textcolor{white}{\scriptsize \sffamily 2008-07-08 02:49:01}}
        \put(8.8,8.36){\textcolor{white}{\scriptsize \sffamily STEREO Ahead HI1}}
        \put(8.8,4.54){\textcolor{white}{\scriptsize \sffamily 2008-07-08 04:49:01}}
        \put(0.96,4.1){\textcolor{white}{\scriptsize \sffamily STEREO Ahead HI1}}
        \put(1.1,0.25){\textcolor{white}{\scriptsize \sffamily 2008-07-08 06:49:01}}
        \put(4.9,4.1){\textcolor{white}{\scriptsize \sffamily STEREO Ahead HI1}}
        \put(4.9,0.25){\textcolor{white}{\scriptsize \sffamily 2008-07-08 08:49:01}}
        \put(8.8,4.1){\textcolor{white}{\scriptsize \sffamily STEREO Ahead HI1}}
        \put(8.8,0.25){\textcolor{white}{\scriptsize \sffamily 2008-07-08 10:49:01}}
        \end{picture}
   \caption{Flux rope CME event and its evolution in the interplanetary medium. Top left panel: the CME on 7 July 2008 in the field of view of the COR2-A coronagraph, with the circular flux rope structure outlined by the green dotted circle. Remaining panels: The same CME on 8 July 2008, seen evolving with time in the field of view of the HI1-A instrument. The light blue dots outline the distorted cavity as it flattens with time, in comparison with a circular cross-section.}
   \label{fig:figureWL}
\end{figure}

Distortions are not only evident in in situ observations, but also they can be distinguished remotely in white light imagery. This is the case of the CME on July 7, 2008 seen in the top left panel of Figure~\ref{fig:figureWL} while crossing the field of view of the STEREO/SECCHI COR2-A coronagraph. Its main axis of symmetry is approximately aligned with STEREO-A's line of sight, so that the cross-section of the magnetic flux rope can be discerned in the image as circular threads outlining the dark circular cavity (indicated by the green dashed circle). The other panels of Figure~\ref{fig:figureWL} show the same CME as it evolves in the interplanetary medium, within the field of view of the STEREO/SECCHI HI1-A telescope. At these distances, the dark circular cavity is distorted into a heart-shaped form (light blue dots) that flattens with distance. This case would be comparable to the one illustrated in Figure~\ref{fig:figure2}d, but considerably flattened. 

\begin{figure}
    \includegraphics[width=0.59\textwidth]{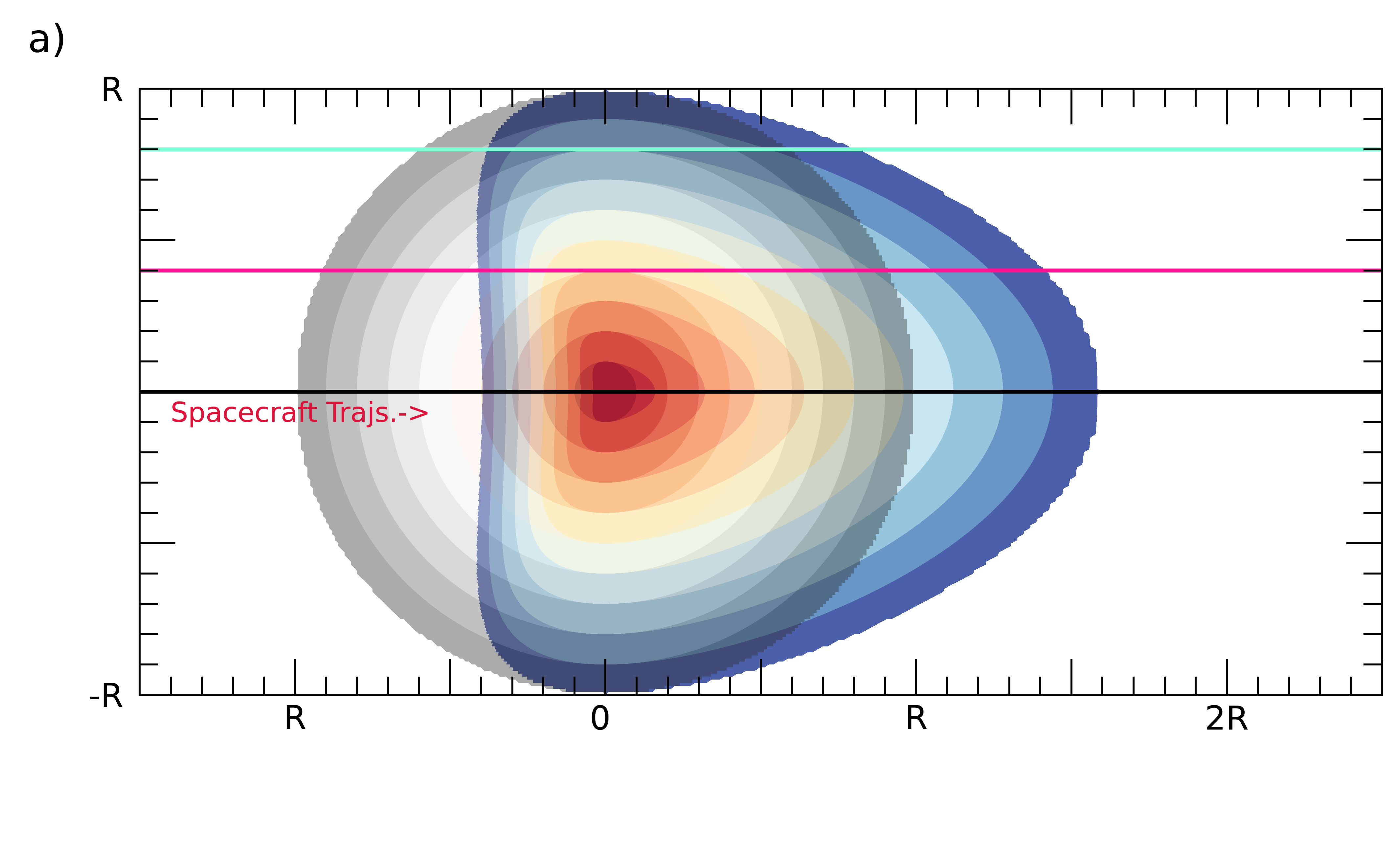} \\
    \includegraphics[width=0.59\textwidth]{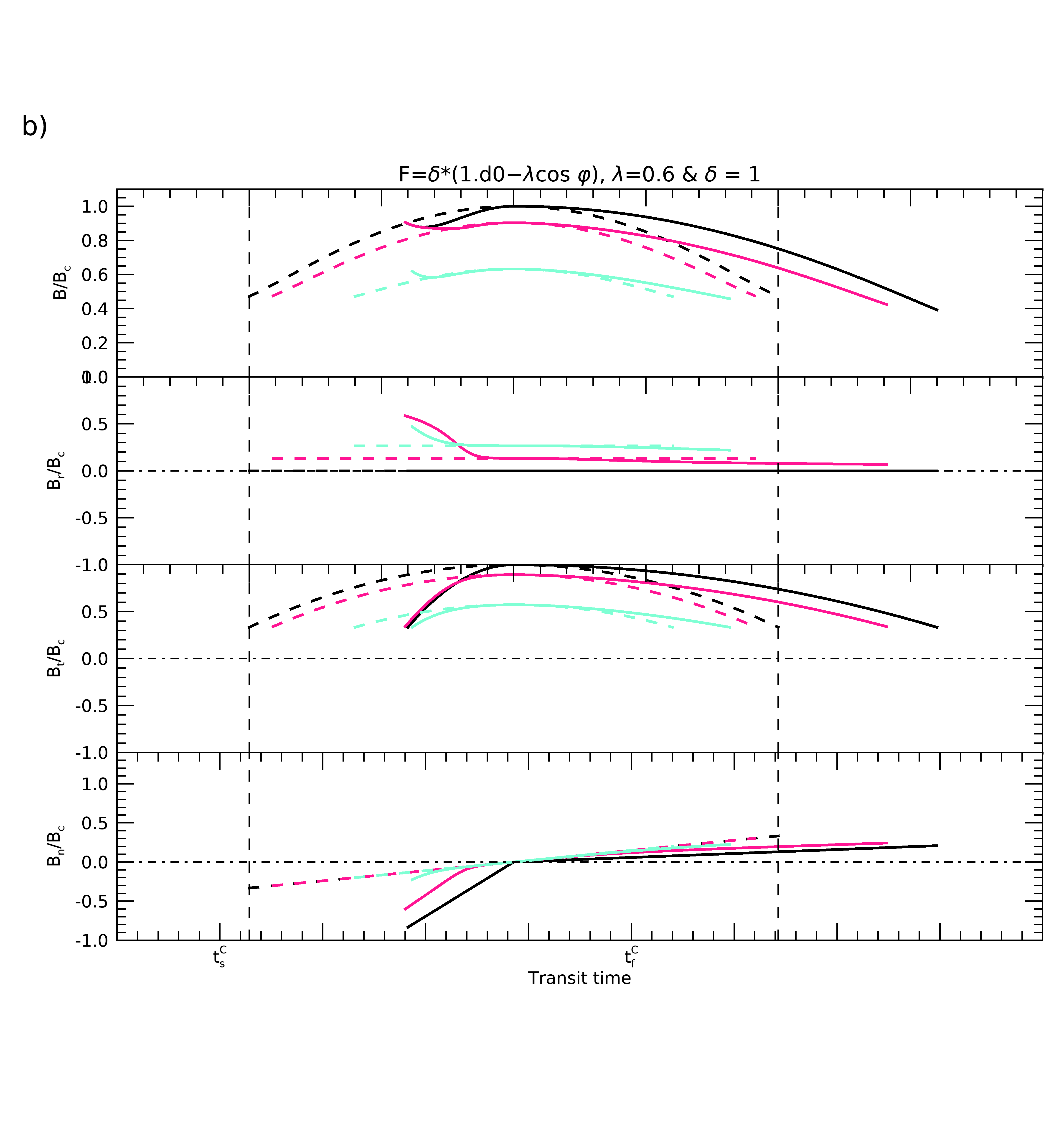}
   \caption{(a) Circular cross-section flux rope (black and white) and overplotted a distorted cross-sections (colored) based on the $F=\delta(1-\lambda\cos\varphi)$ geometry. The three colored (black, red, light blue) lines indicate a spacecraft trajectory crossing from the left. (b) Simulation of the expected in situ observations from a spacecraft crossing a distorted and a non-distorted flux-rope at different distance from the center.  }
   \label{fig:figure7}
\end{figure}

Figure~\ref{fig:figure7} illustrates the impact of the distortion in one of the critical elements for a reliable Space Weather capability, the arrival time. At the top, Figure~\ref{fig:figure7}(a) displays in black and white a circular cross-section and a colored distorted cross-section based on the function $F=\delta(1-\lambda\cos\varphi)$. This overlay image could represent a case of a distorted structure fitted by the conventional Gradual Cylindrical Shell (GCS) forward-modeling technique \citep[][]{Thernisien2006,Thernisien2009}, which is only capable of simulating structures with circular cross-sections. This could be one of the options in the attempt to fit the GCS technique in the actual observations, very similar to the illustration in the Figure~\ref{fig:figureWL}. The Figure~\ref{fig:figure7}(a) includes a three spacecraft crossing with trajectories from the left through the center (black line), half way to the edge (red line) and very close to the edge (light blue line). 
Figure~\ref{fig:figure7}(b) includes a comparative analysis of the simulated in situ  magnetic field observations of a spacecraft crossing a circular (dashed line) and a distorted structure (solid line). This exercise illustrates the probably cause of the off in the prediction of the CME arrival time. By assuming that the GCS fits the if we would like to make a prediction of the arrival time on the basis of the non-distorted model, the error will be, at least 1/3 of the crossing duration. The peak or maximum in the magnetic field will remain in the same location since the compression does not displace the flux rope center and the duration will be very similar. The location of the sign change crossing in the $B_N$ polarity won't change but the duration with negative polarity will be, at least, half shorter than expected, and the $B_N$ positive polarity duration will be larger than predicted.

Physical parameters, such as the magnetic fluxes, are also relevant for space weather studies. In the case of the poloidal flux, the impact of the distortion on this  parameter would depend on the accurate of the fitting either based on remote-sensing or in situ observations. Thus, based on the Equation~(\ref{Eq:Flujo_varphi}), the poloidal magnetic flux is not impacted by the distortion if the estimation of the radial size ($R$) of the structure is right. In the case of the toroidal magnetic flux, the deviation from the actual value will depend on the integral,
\begin{equation}
    \int_0^{2\pi} \overline{g} d\varphi.
\end{equation}

This factor will change the expected $2\pi$ value for the CC case in general in the case of the axial or toroidal magnetic flux. For instance, in the elliptical case, Figure~\ref{fig:figure2}(a), the result of the integral will reduce the toroidal magnetic flux in a $\delta$ value of the expected or assumed CC case. This will be also the effect in the case-example that we have developed in this paper with $F(\varphi)=\delta(1-\lambda\cos\varphi)$. However, in this last case, there will be an effect in the calculation of the accumulative magnetic flux at the compressed area that may drive to interpret the distortion as erosion effects \citep[see for instance, ][]{Dasso2006,Pal2021}.

\section{Summary and Final Remarks}

This paper presents an improvement to the circular-cilindrical (CC) and elliptic-cylindrical (EC) models \citep[][]{Nieves-Chinchilla2016,Nieves-Chinchilla2018ApJ}. Based on the mathematical formulation developed for the EC model, we have moved forward to more complex and perhaps more realistic distortions of heliospheric flux ropes. 
Here we have presented the model based on a toroidal geometry and expanded to a general cross-section distortion. We have illustrated the model with four functions, the morphology and internal magnetic field distribution in the Section~\ref{Sec:CoordSys}. As a proof of concept, in the Section~\ref{sec:section3}, we have developed the model to a particular cross section based on the geometry $F=\delta(1-\lambda\cos\varphi)$. 

We have now incorporated a whole section (Section~\ref{sec:section3}) to understand the in situ implications of the distortion on the heliospheric flux ropes. We have simulated two crossing of a spacecraft through a flux rope (center and halfway to the outer boundary) with two different curvatures and compared with a regular CC geometry.
We have mapped (contour plot) the magnetic field magnitude and the three RTN magnetic components for the three cross-sections and discussed the changes in the profiles of the magnitude and components that the magnetometer would record in the two crossings and compared with the CC for two different global curvatures. 

The exercise carried out in section~\ref{sec:section3} provides the opportunity to train our `brain' and look for examples in the real observations. In Section~\ref{Sec:section4} we have included two real events from the Wind ICME catalog \citep{Nieves-Chinchilla2018SolPhy} with magnetic configurations that depart from the expected CC geometry. 

In this paper, we argue that by doing the fitting of any of the events included in this paper and by increasing the number of parameters, we may find a good fitting but we won’t be able to ensure that they are the right ones. 
This is an exercise carried out for many years, not only by the authors of this paper but also by many others. The reconciliation with the multiple observations, including imaging, and new numerical techniques is far to be successful \citep[see for instance, ][]{Al-haddad2013,Woodetal2017}. This is why we think that, in order to develop a robust 3D physic-driven modeling and reconstruction technique, the in situ fitting procedure by itself is not enough to reproduce the physical and morphological characteristics of the flux ropes. The inclusion of the imaging observations in the reconstructions techniques, accompanying by a brain- and machine- training techniques, and eventually the evaluation of dynamics associated to the distortion and deformation \citep[][]{Kay&Nieves2021a,Kay&Nieves2021b}, are the elements to take into account before confirming this or any other other model is the right one to describe the flux ropes in the heliosphere





\begin{acknowledgments}
The work of TNC is s supported by the National Aeronautics and Space Administration under Grant 80NSSC19K0274 issued through the Heliophysics Guest Investigators Program and the Solar Orbiter mission.
HC is member of the ``Carrera del Investigador Cient\'ifico" of CONICET. We acknowledge use of data from the STEREO (NASA) mission, produced by the SECCHI consortium. The displayed STEREO/SECCHI HI-1 images are a Level-2 product from the Southwest Research Institute in Boulder, Colorado.
\end{acknowledgments}

%

\vspace{5mm}
\facilities{}

\bibliography{sample631.bib}

\begin{thebibliography}{}
\expandafter\ifx\csname natexlab\endcsname\relax\def\natexlab#1{#1}\fi
\providecommand{\url}[1]{\href{#1}{#1}}
\providecommand{\dodoi}[1]{doi:~\href{http://doi.org/#1}{\nolinkurl{#1}}}
\providecommand{\doeprint}[1]{\href{http://ascl.net/#1}{\nolinkurl{http://ascl.net/#1}}}
\providecommand{\doarXiv}[1]{\href{https://arxiv.org/abs/#1}{\nolinkurl{https://arxiv.org/abs/#1}}}

\bibitem[{{Al-Haddad} {et~al.}(2013){Al-Haddad}, {Nieves-Chinchilla}, {Savani},
  {M{\"o}stl}, {Marubashi}, {Hidalgo}, {Roussev}, {Poedts}, \&
  {Farrugia}}]{Al-haddad2013}
{Al-Haddad}, N., {Nieves-Chinchilla}, T., {Savani}, N.~P., {et~al.} 2013,
  \solphys, 284, 129, \dodoi{10.1007/s11207-013-0244-5}

\bibitem[{{Arfken} \& {Weber}(2005)}]{Arfken2005}
{Arfken}, G.~B., \& {Weber}, H.~J. 2005, {Mathematical methods for physicists
  6th ed.}

\bibitem[{{Brown} {et~al.}(1999){Brown}, {Canfield}, \& {Pevtsov}}]{Brown1999}
{Brown}, M.~R., {Canfield}, R.~C., \& {Pevtsov}, A.~A. 1999, Washington DC
  American Geophysical Union Geophysical Monograph Series, 111,
  \dodoi{10.1029/GM111}

\bibitem[{{Burlaga} {et~al.}(1981){Burlaga}, {Sittler}, {Mariani}, \&
  {Schwenn}}]{Burlaga1981}
{Burlaga}, L., {Sittler}, E., {Mariani}, F., \& {Schwenn}, R. 1981, \jgr, 86,
  6673, \dodoi{10.1029/JA086iA08p06673}

\bibitem[{{Dasso} {et~al.}(2006){Dasso}, {Mandrini}, {D{\'e}moulin}, \&
  {Luoni}}]{Dasso2006}
{Dasso}, S., {Mandrini}, C.~H., {D{\'e}moulin}, P., \& {Luoni}, M.~L. 2006,
  \aap, 455, 349, \dodoi{10.1051/0004-6361:20064806}

\bibitem[{{D{\'e}moulin} {et~al.}(2013){D{\'e}moulin}, {Dasso}, \&
  {Janvier}}]{Demoulin2013}
{D{\'e}moulin}, P., {Dasso}, S., \& {Janvier}, M. 2013, \aap, 550, A3,
  \dodoi{10.1051/0004-6361/201220535}

\bibitem[{{Domingo} {et~al.}(1995){Domingo}, {Fleck}, \&
  {Poland}}]{domingo1995}
{Domingo}, V., {Fleck}, B., \& {Poland}, A.~I. 1995, Solar Physics, 162, 1,
  \dodoi{10.1007/BF00733425}

\bibitem[{{dos Santos} {et~al.}(2020){dos Santos}, {Narock},
  {Nieves-Chinchilla}, {Nu{\~n}ez}, \& {Kirk}}]{dosSantos2020}
{dos Santos}, L. F.~G., {Narock}, A., {Nieves-Chinchilla}, T., {Nu{\~n}ez}, M.,
  \& {Kirk}, M. 2020, \solphys, 295, 131, \dodoi{10.1007/s11207-020-01697-x}

\bibitem[{{Fox} {et~al.}(2016){Fox}, {Velli}, {Bale}, {Decker}, {Driesman},
  {Howard}, {Kasper}, {Kinnison}, {Kusterer}, {Lario}, {Lockwood}, {McComas},
  {Raouafi}, \& {Szabo}}]{Fox2016}
{Fox}, N.~J., {Velli}, M.~C., {Bale}, S.~D., {et~al.} 2016, \ssr, 204, 7,
  \dodoi{10.1007/s11214-015-0211-6}

\bibitem[{{Hau} \& {Sonnerup}(1999)}]{HauSonnerup1999}
{Hau}, L.~N., \& {Sonnerup}, B. U.~{\"O}. 1999, \jgr, 104, 6899,
  \dodoi{10.1029/1999JA900002}

\bibitem[{{Hidalgo} {et~al.}(2002{\natexlab{a}}){Hidalgo}, {Cid}, {Vinas}, \&
  {Sequeiros}}]{Hidalgo2002}
{Hidalgo}, M.~A., {Cid}, C., {Vinas}, A.~F., \& {Sequeiros}, J.
  2002{\natexlab{a}}, \jgr, 107, 1002, \dodoi{10.1029/2001JA900100}

\bibitem[{{Hidalgo} {et~al.}(2002{\natexlab{b}}){Hidalgo}, {Nieves-Chinchilla},
  \& {Cid}}]{Hidalgo2002a}
{Hidalgo}, M.~A., {Nieves-Chinchilla}, T., \& {Cid}, C. 2002{\natexlab{b}},
  \grl, 29, 130000, \dodoi{10.1029/2001GL013875}

\bibitem[{{Hu}(2017)}]{Hu2017}
{Hu}, S.~Q. 2017, Science China Earth Sciences, 60, 1466,
  \dodoi{10.1007/s11430-017-9067-2}

\bibitem[{{Kaiser} {et~al.}(2008){Kaiser}, {Kucera}, {Davila}, {St.~Cyr},
  {Guhathakurta}, \& {Christian}}]{Kaiser2008}
{Kaiser}, M.~L., {Kucera}, T.~A., {Davila}, J.~M., {et~al.} 2008, \ssr, 136, 5,
  \dodoi{10.1007/s11214-007-9277-0}

\bibitem[{{Kay} \& {Nieves-Chinchilla}(2021{\natexlab{a}})}]{Kay&Nieves2021a}
{Kay}, C., \& {Nieves-Chinchilla}, T. 2021{\natexlab{a}}, Journal of
  Geophysical Research (Space Physics), 126, 2020JA028911,
  \dodoi{10.1029/2020JA028911}

\bibitem[{{Kay} \& {Nieves-Chinchilla}(2021{\natexlab{b}})}]{Kay&Nieves2021b}
---. 2021{\natexlab{b}}, Journal of Geophysical Research (Space Physics), 126,
  2020JA028911, \dodoi{10.1029/2020JA028911}

\bibitem[{{Kilpua} {et~al.}(2017){Kilpua}, {Koskinen}, \&
  {Pulkkinen}}]{Kilpua2017}
{Kilpua}, E., {Koskinen}, H. E.~J., \& {Pulkkinen}, T.~I. 2017, Living Reviews
  in Solar Physics, 14, 5, \dodoi{10.1007/s41116-017-0009-6}

\bibitem[{{Lepping} {et~al.}(1990){Lepping}, {Burlaga}, \&
  {Jones}}]{Lepping1990}
{Lepping}, R.~P., {Burlaga}, L.~F., \& {Jones}, J.~A. 1990, \jgr, 95, 11957,
  \dodoi{10.1029/JA095iA08p11957}

\bibitem[{{Lundquist}(1951)}]{Lundquist1951}
{Lundquist}, S. 1951, Physical Review, 83, 307, \dodoi{10.1103/PhysRev.83.307}

\bibitem[{{M{\"u}ller} {et~al.}(2020){M{\"u}ller}, {St. Cyr}, {Zouganelis},
  {Gilbert}, {Marsden}, {Nieves-Chinchilla}, {Antonucci}, {Auch{\`e}re},
  {Berghmans}, {Horbury}, {Howard}, {Krucker}, {Maksimovic}, {Owen}, {Rochus},
  {Rodriguez-Pacheco}, {Romoli}, {Solanki}, {Bruno}, {Carlsson}, {Fludra},
  {Harra}, {Hassler}, {Livi}, {Louarn}, {Peter}, {Sch{\"u}hle}, {Teriaca}, {del
  Toro Iniesta}, {Wimmer-Schweingruber}, {Marsch}, {Velli}, {De Groof},
  {Walsh}, \& {Williams}}]{Muller2020}
{M{\"u}ller}, D., {St. Cyr}, O.~C., {Zouganelis}, I., {et~al.} 2020, \aap, 642,
  A1, \dodoi{10.1051/0004-6361/202038467}

\bibitem[{{Narock} {et~al.}(2022){Narock}, {Narock}, {Dos Santos}, \&
  {Nieves-Chinchilla}}]{Narock2022}
{Narock}, T., {Narock}, A., {Dos Santos}, L. F.~G., \& {Nieves-Chinchilla}, T.
  2022, Frontiers in Astronomy and Space Sciences, 9, 838442,
  \dodoi{10.3389/fspas.2022.838442}

\bibitem[{{Nieves-Chinchilla} {et~al.}(2018){Nieves-Chinchilla}, {Linton},
  {Hidalgo}, \& {Vourlidas}}]{Nieves-Chinchilla2018ApJ}
{Nieves-Chinchilla}, T., {Linton}, M.~G., {Hidalgo}, M.~A., \& {Vourlidas}, A.
  2018, \apj, 861, 139, \dodoi{10.3847/1538-4357/aac951}

\bibitem[{Nieves-Chinchilla {et~al.}(2016)Nieves-Chinchilla, Linton, Hidalgo,
  Vourlidas, Savani, Szabo, Farrugia, \& Yu}]{Nieves-Chinchilla2016}
Nieves-Chinchilla, T., Linton, M.~G., Hidalgo, M.~A., {et~al.} 2016, The
  Astrophysical Journal, 823, 27.
\newblock \url{http://stacks.iop.org/0004-637X/823/i=1/a=27}

\bibitem[{Nieves-Chinchilla {et~al.}(2018)Nieves-Chinchilla, Vourlidas,
  Raymond, Linton, Al-haddad, Savani, Szabo, \&
  Hidalgo}]{Nieves-Chinchilla2018SolPhy}
Nieves-Chinchilla, T., Vourlidas, A., Raymond, J.~C., {et~al.} 2018, Solar
  Physics, 293, 25, \dodoi{10.1007/s11207-018-1247-z}

\bibitem[{{Pal} {et~al.}(2021){Pal}, {Kilpua}, {Good}, {Pomoell}, \&
  {Price}}]{Pal2021}
{Pal}, S., {Kilpua}, E., {Good}, S., {Pomoell}, J., \& {Price}, D.~J. 2021,
  \aap, 650, A176, \dodoi{10.1051/0004-6361/202040070}

\bibitem[{{Pesnell} {et~al.}(2012){Pesnell}, {Thompson}, \&
  {Chamberlin}}]{Pesnell-etal2012}
{Pesnell}, W.~D., {Thompson}, B.~J., \& {Chamberlin}, P.~C. 2012, Solar
  Physics, 275, 3, \dodoi{10.1007/s11207-011-9841-3}

\bibitem[{{Rouillard}(2011)}]{Rouillard2011}
{Rouillard}, A.~P. 2011, Journal of Atmospheric and Solar-Terrestrial Physics,
  73, 1201, \dodoi{10.1016/j.jastp.2010.08.015}

\bibitem[{{Sonnerup} \& {Guo}(1996)}]{Sonnerup1996}
{Sonnerup}, B.~U.~{\"O}., \& {Guo}, M. 1996, \grl, 23, 3679,
  \dodoi{10.1029/96GL03573}

\bibitem[{{Suess}(1988)}]{Suess1988}
{Suess}, S.~T. 1988, \jgr, 93, 5437, \dodoi{10.1029/JA093iA06p05437}

\bibitem[{{Taylor}(1974)}]{Taylor1974}
{Taylor}, J.~B. 1974, \prl, 33, 1139, \dodoi{10.1103/PhysRevLett.33.1139}

\bibitem[{{Thernisien} {et~al.}(2009){Thernisien}, {Vourlidas}, \&
  {Howard}}]{Thernisien2009}
{Thernisien}, A., {Vourlidas}, A., \& {Howard}, R.~A. 2009, \solphys, 256, 111,
  \dodoi{10.1007/s11207-009-9346-5}

\bibitem[{{Thernisien} {et~al.}(2006){Thernisien}, {Howard}, \&
  {Vourlidas}}]{Thernisien2006}
{Thernisien}, A.~F.~R., {Howard}, R.~A., \& {Vourlidas}, A. 2006, \apj, 652,
  763, \dodoi{10.1086/508254}

\bibitem[{{Woltjer}(1958)}]{Woltjer1958}
{Woltjer}, L. 1958, Proceedings of the National Academy of Science, 44, 489,
  \dodoi{10.1073/pnas.44.6.489}

\bibitem[{{Wood} {et~al.}(2017){Wood}, {Wu}, {Lepping}, {Nieves-Chinchilla},
  {Howard}, {Linton}, \& {Socker}}]{Woodetal2017}
{Wood}, B.~E., {Wu}, C.-C., {Lepping}, R.~P., {et~al.} 2017, \apjs, 229, 29,
  \dodoi{10.3847/1538-4365/229/2/29}

\end{thebibliography}
\bibliographystyle{aasjournal}



\end{document}